\documentclass{iopart}
\usepackage{graphicx}
\usepackage{epstopdf}

\begin{document}
\title[Dynamics of artificial spin ice]
{Dynamics of artificial spin ice: continuous honeycomb network}
\author{Yichen Shen$^{1}$}
\author{Olga Petrova$^2$}
\author{Paula Mellado$^3$}
\author{Stephen Daunheimer$^4$}
\author{John Cumings$^4$}
\author{Oleg Tchernyshyov$^2$}
\address{$^1$Department of Physics, Massachusetts Institute of Technology, Cambridge, Massachusetts 02139, USA}
\address{$^2$Department of Physics and Astronomy, Johns Hopkins University, Baltimore, Maryland 21218, USA}
\address{$^3$School of Engineering and Applied Sciences, Harvard University, Cambridge, Massachusetts 02138, USA}
\address{$^4$Department of Materials Science and Engineering, University of Maryland, College Park, 20742 Maryland, USA}

\ead{cumings@umd.edu, olegt@jhu.edu}

\begin{abstract}
We model the dynamics of magnetization in an artificial analog of spin ice specializing to the case of a honeycomb network of connected magnetic nanowires. The inherently dissipative dynamics is mediated by the emission, propagation and absorption of domain walls in the links of the lattice. These domain walls carry two natural units of magnetic charge, whereas sites of the lattice contain a unit magnetic charge. Magnetostatic Coulomb forces between these charges play a major role in the physics of the system, as does quenched disorder caused by imperfections of the lattice.  We identify and describe different regimes of magnetization reversal in an applied magnetic field determined by the orientation of the applied field with respect to the initial magnetization. One of the regimes is characterized by magnetic avalanches with a $1/n$ distribution of lengths. 
\end{abstract}

\section{Introduction}
\label{sec:intro}

Spin ice \cite{PhysRevLett.79.2554, gingras.review.2009} is a frustrated ferromagnet with Ising spins that possesses rather peculiar properties.  First, as a consequence of strong frustration, it has a massively degenerate ground state and retains a finite entropy density even at very low temperatures \cite{nature.399.333}. Second, its low-energy excitations are neither individual flipped spins, nor domain walls, but are point defects acting as sources and sinks of magnetic field $\mathbf H$ \cite{JETP.101.481, Nature.451.42}. The concept of magnetic charges, while not exactly new \cite{LL8.44, Jackson.5.9, saitoh:2004}, has proven very useful in elucidating the static and dynamic properties of spin ice \cite{NatPhys.5.258, Science.326.411, Nature.461.956, Science.326.415, JPSJ.78.103706}. It is worth noting that these objects are magnetic analogs of excitations with fractional electric charge found in the familiar water ice \cite{petrenko.book}. 

Artificial spin ice is an array of nanomagnets with similarly frustrated interactions. The original system made by Schiffer's group had disconnected elongated islands (80 nm by 220 nm laterally and 25-nm thick) made of permalloy and arranged as links of a square lattice \cite{nature.439.303}. Later versions included a connected honeycomb network of flat magnetic wires \cite{tanaka:052411, qi:094418, Ladak:2010, NJP.13.023023}, in which the centers of the wires form a kagome lattice, hence the sometimes used name ``kagome spin ice'' \cite{qi:094418}. Whereas it had been originally intended as a large-scale replica of natural spin ice, it became clear very soon that artificial spin ice has a number of its own peculiar features. For example, because the magnetic moments in artificial spin ice are extremely large, on the order of $10^8$ Bohr magnetons, the energy scale of shape anisotropy due to dipolar interactions, $10^5$ K in temperature units \cite{PhysRevLett.98.217203}, effectively freezes out thermal fluctuations of the macrospins meaning that the system is not in thermal equilibrium. Dynamics of magnetization has to be induced by the application of an external magnetic field \cite{nature.439.303}. Elaborate experimental protocols involving a magnetic field of varying magnitude and direction \cite{JApplPhys.101.09J104} have been proposed to simulate thermal agitation invoking parallels with fluidized granular matter.  It remains to be seen whether the induced dynamics yields a thermal ensemble with an effective temperature.  The analogy with granular matter is further reinforced by recent observations of magnetic avalanches in the process of magnetization reversal \cite{Ladak:2010, NatPhys.7.68}. 

In this paper we present a model of magnetization dynamics in artificial spin ice subject to an external magnetic field. Two sets of physical variables are used: an Ising variable $\sigma = \pm 1$ encodes the magnetic state of a spin, whereas an integer $q$ quantifies the magnetic charge of a node at the junction of several spins. Magnetization dynamics are mediated by the emission of domain walls carrying two units of magnetic charge from a lattice node, their subsequent propagation through a magnetic element, and absorption at the next node.  We specialize to the case of kagome spin ice, in which magnetic elements form a connected honeycomb lattice \cite{Ladak:2010, tanaka:052411, qi:094418, NJP.13.023023}. The model can be readily extended to other geometries and lattices with disconnected magnetic elements \cite{nature.439.303, NatPhys.7.68, NatPhys.7.75, PhysRevLett.106.057209}. Some of the results presented here have been outlined previously \cite{PRL.105.187206}.

\section{Basic features of the model}
\label{sec:basics}

Our model is specialized toward an experimental realization described previously \cite{qi:094418}. That artificial spin ice is a connected honeycomb network of permalloy nanowires with saturation magnetization $M = 8.6 \times 10^5$ A/m and the following typical dimensions: length $l = 500$ nm, width $w = 110$ nm, and thickness $t = 23$ nm.  Three nanowires come together at a vertex in the bulk.  At the edge of the lattice, a vertex may have one or two links coming in. 

\subsection{Basic variables: magnetization and magnetic charge}
\label{sec:basics-variables}

We label nodes of the lattice by a single index $i$ and nanowires connecting adjacent nodes by the indices of its two nodes, $ij$. In equilibrium, the vector of magnetization $\mathbf M$ points parallel to the long axis of the wire, so we can encode the two states of a nanowire by using an Ising variable $\sigma_{ij} = \pm 1$.  In our convention, $\sigma_{ij} = +1$ when the vector of magnetization points from node $i$ into node $j$. This definition implies antisymmetry under index exchange, $\sigma_{ij} = -\sigma_{ji}$.

\begin{figure}
\begin{center}
\includegraphics[width=0.5\columnwidth]{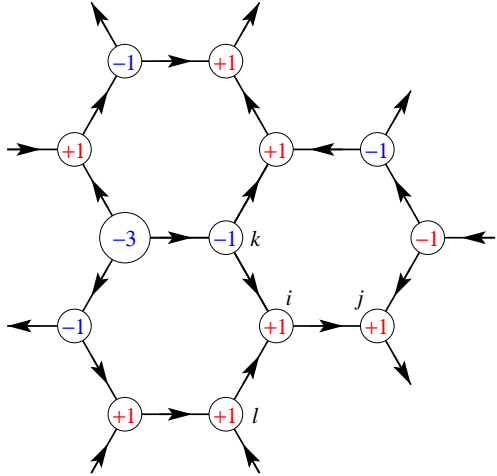}
\end{center}
\caption{Magnetization variables $\sigma_{ij} = \pm 1$ (arrows) live on links $ij$ of the honeycomb lattice.  Charges $q_i = \pm 1, \, \pm 3$ live on nodes $i$.}
\label{fig:variables}
\end{figure}

We define the dimensionless magnetic charge at node $i$ as 
\begin{equation}
q_i = \sum_{j} \sigma_{ji},
\label{eq:q-def}
\end{equation}
where the sum is taken over the three neighboring sites $j$.  This definition is quite natural: since magnetic induction $\mathbf B = \mu_0(\mathbf H + \mathbf M)$ is divergence-free, the magnetic charge $Q_i$ of node $i$ equals the flux of magnetic field $\mathbf H$ out of the node, which in turn equals the flux of magnetization $\mathbf M$ into it:
\begin{equation}
Q_i = \oint \mathbf H \cdot \rmd \mathbf A 
	= -\oint \mathbf M \cdot \rmd \mathbf A 
	= - Mtw \sum_{j} \sigma_{ij}
	= Mtw q_i.
\end{equation}
Thus $q_i$ is indeed magnetic charge measured in units of $Mtw$. 

The Bernal-Fowler ice rule \cite{gingras.review.2009} enforcing minimization of the absolute value of charge $|Q_i|$ is usually justified from the energy perspective: the magnetostatic energy of spin ice can be written as the energy of Coulomb interaction of magnetic charges, 
\begin{equation}
E \approx \frac{\mu_0}{8\pi}\sum_{i \neq j}\frac{Q_i Q_j}{|\mathbf r_i - \mathbf r_j|}
+ \sum_{i} \frac{Q_i^2}{2C}.
\end{equation}
The dominant second term---the charging energy of a node---forces minimization of magnetic charges in natural spin ice. The ``capacitance'' $C$ is determined by the dipolar and exchange couplings energies of adjacent spins \cite{Nature.451.42}. 

Although we will see below that these energy considerations are not relevant to artificial spin ice within our model, for the moment we will simply adopt the result to it. In honeycomb ice, where the coordination number is 3, dimensionless charge $q_i$ can take on values $\pm 1$ and $\pm 3$.  Minimization of node self-energy would select states with 
\begin{equation}
q_i = \pm 1.  
\label{eq:ice-rule}
\end{equation}
Indeed, triple magnetic charges have never been observed in our samples of honeycomb ice.  Ladak \textit{et al.} \cite{Ladak:2010, NJP.13.023023} have found nodes with triple charges. The difference is likely due to a higher amount of quenched disorder arising from random imperfections of the lattice \cite{PhysRevLett.107.217204} in the samples of Ladak \textit{et al.} 

We will find it convenient to use the following notation. A site with a unit charge $q_i = \pm 1$ has two \textit{majority} links with $\sigma_{ji} = q_i$ and one \textit{minority} link with $\sigma_{ji} = -q_i$.  For site $i$ in Figure~\ref{fig:variables}, the minority link is $ij$.

\subsection{Basic dynamics: emission of a domain wall}
\label{sec:basics-emission}

\begin{figure}
\begin{center}
\includegraphics[width=0.25\columnwidth]{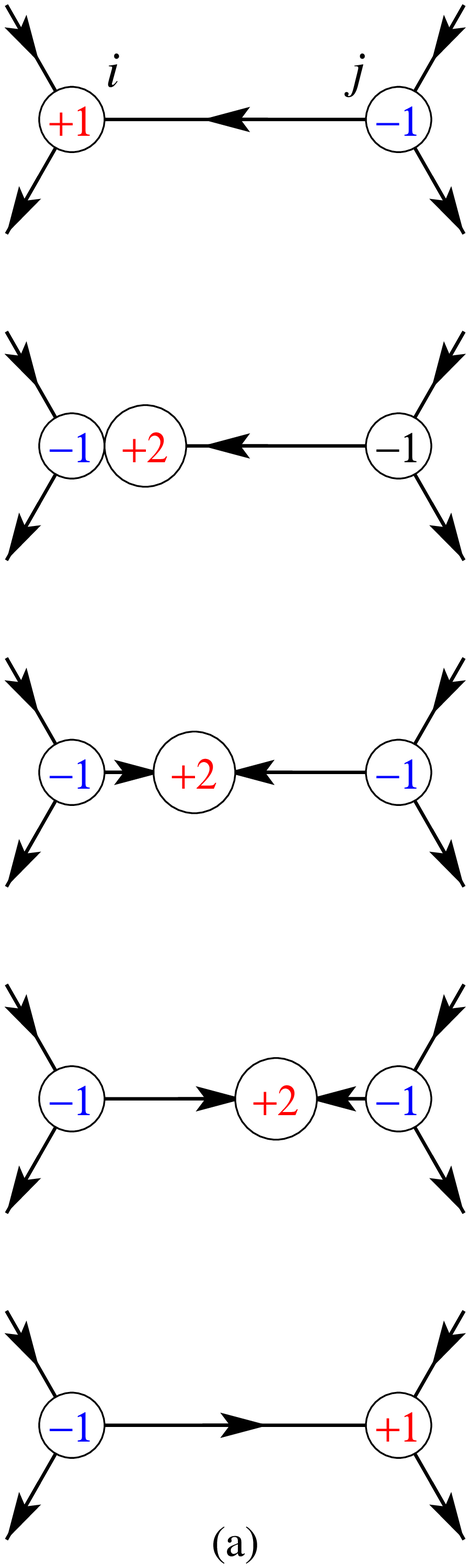}
\hskip 10mm
\includegraphics[width=0.258\columnwidth]{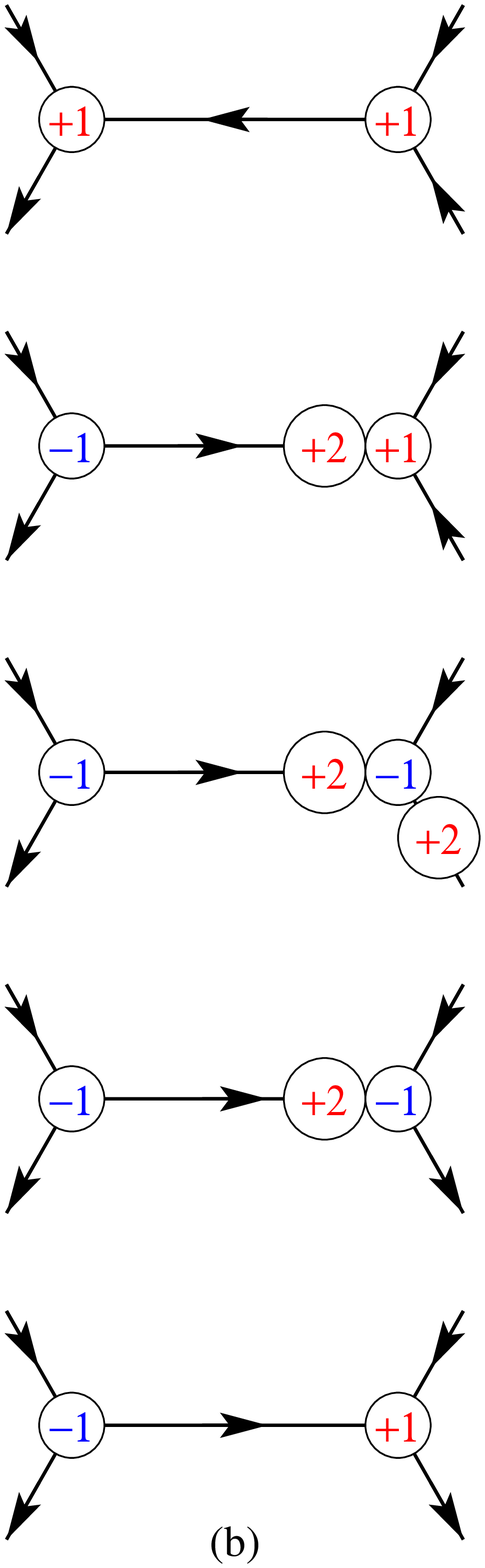}
\end{center}
\caption{Magnetization reversal in a single link. At the end of the reversal, the domain wall encounters a node with magnetic charge of the opposite sign (a) or of the same sign (b). In panel (b), the emission of the domain wall from the left node and its propagation along the horizontal link are omitted for brevity.}
\label{fig:reversal-link}
\end{figure}

To reverse the magnetization in a nanowire, one must apply a sufficiently strong external magnetic field. The reversal begins when one of the nodes, say $i$, emits a domain wall (w) into link $ij$, Figure \ref{fig:reversal-link}(a).  If the link initially has magnetization $\sigma_{ij} = \pm 1$, a domain wall can traverse it from $i$ to $j$ only if it has charge of the right sign, i.e., $q_\mathrm{w} = 2 \sigma_{ij} = \pm 2$.  Once the domain wall passes through the link, $\sigma_{ij}$ changes its sign. Now a domain wall with the same charge $q_\mathrm{w}$ can only traverse the link in the opposite direction. 

The critical field $H_\mathrm{c}$, at which a domain wall is emitted from a node, can be estimated as follows \cite{NatPhys.6.323}.  Suppose a node with magnetic charge $q_i = \pm 1$ emits a domain wall with magnetic charge $q_\mathrm{w} = \pm 2$ \cite{saitoh:2004, PRL.105.187206}.  Conservation of magnetic charge means that the charge of the site turns to $q_i = \mp 1$. The emission process can thus be viewed as pulling a charge $q_\mathrm{w} = \pm 2$ away from a charge of the opposite sign $q_i = \mp 1$.  The maximum force between the two charges is achieved when the separation between them is of the order of their sizes $a$, which is roughly equal to the width of the wire $w$: $F_\mathrm{max} = \mu_0 |Q_i Q_\mathrm{w}|/(4\pi a^2)$. This force must be overcome by the Zeeman force applied to the domain wall by the external magnetic field, $F_\mathrm{ext} = \mu_0 |Q_\mathrm{w}| H_\mathrm{ext}$. Hence the estimate of the critical field, 
\begin{equation}
H_\mathrm{c} = \frac{|Q_i|}{4\pi a^2} = \frac{Mtw}{4 \pi a^2} \approx \frac{Mt}{4 \pi w}.
\label{eq:Hc-estimate}
\end{equation}
For the system parameters used in our previous work \cite{qi:094418} and listed above, this estimate yields $\mu_0 H_\mathrm{c} = 18$ mT.  The critical value observed experimentally \cite{PhysRevLett.107.167201} is 35 mT.  

One can envision another possible process, wherein the reversal is triggered when a site with charge $q_i = \pm 1$ emits a domain wall of charge $q_\mathrm{w} = \mp 2$ and change its charge to $q_i = \pm 3$.  Considerations along the same lines as above show that the critical field required to pull apart charges $q_i = \pm 3$ and $q_\mathrm{w} = \mp 2$ is $3H_\mathrm{c}$.  As we will see below, magnetization reversal in samples with low quenched disorder occurs well before the external field has a chance to reach this value. This explains why triple charges are never generated as a result of the emission of a domain wall.  

The estimate for the critical field was obtained under the assumption that the external magnetic field $\mathbf H_\mathrm{ext}$ is applied along the link into which the domain wall is emitted. When the field makes angle $\theta$ with the link, it is reasonable to suppose that only the longitudinal component of the field $H_\mathrm{ext} \cos{\theta}$ pulls the domain wall away from the node. We thus expect the following angular dependence of the critical field:
\begin{equation}
H_\mathrm{c}(\theta) = H_\mathrm{c}/\cos{\theta}.
\label{eq:Hc-theta-estimate}
\end{equation}
As we will see later in Sec.~\ref{sec:micro}, our educated guess is almost right and that Eq.~(\ref{eq:Hc-theta-estimate}) requires only a minor correction: the angle $\theta$ should be measured not from the axis of the link but from a slightly offset direction. This effect is caused by an asymmetric distribution of magnetization around a node, which was missed by the simplified, mesoscopic model of this section.

\subsection{Basic physics: absorption of a domain wall}
\label{sec:basics-absorption}

Once a domain wall is emitted into link $ij$, it quickly propagates to the other end of the link, toward node $j$. Theoretical and experimental studies of domain wall motion in permalloy nanowires \cite{ThiavilleBook06, Atkinson2006} show that walls move at speeds of the order of 100 m/s in an applied field of just 1 mT. This corresponds to a propagation time of the order of 10 ns, which is too short to be observed in most current experimental setups.  

When the domain wall reaches the opposite end of the link $ij$, its further fate depends on whether the magnetic charge at node $j$ has the same or opposite sign of magnetic charge. We consider the two cases in turn. 

If the domain wall and node $j$ at which it arrives have opposite charges, $q_\mathrm{w} = \pm2 = -2 q_j$, as in Figure \ref{fig:reversal-link}(a), the domain wall is attracted to the node. It is easily absorbed by the node, whose charge changes to $q_j = \pm 1$.  A new domain wall with the same charge $q_\mathrm{w} = \pm 2$ may be subsequently emitted into one of the adjacent links $jk$ if two conditions are met: (i) the link has the right direction of its magnetization, $q_\mathrm{w} = 2\sigma_{jk}$ and (ii) the external field is sufficiently strong to trigger the emission. 

Note that condition (ii) is sensitive to the orientation of the field relative to link $jk$.  It also rests on an implicit assumption that the critical field for a new domain wall is not affected by the just completed absorption of the previous one. This assumption is reasonable if the dynamics of domain walls are strongly dissipative and the energy generated during the absorption process is quickly dissipated as heat.  Experiments with domain walls in nanowires indicate that they possess non-negligible inertia \cite{saitoh:2004}, and therefore our assumption of strongly overdamped dynamics may not be fully justified. Nonetheless, for the sake of simplicity, we shall assume that the dynamics are strongly dissipative and that the extra energy brought by the arrival of a domain wall does not by itself cause the emission of another domain wall from the same node.  

Consider now the other case, where the domain wall and the arrival node have charges of the same sign, $q_\mathrm{w} = \pm2 = 2 q_j$, as in Figure \ref{fig:reversal-link}(b). The two charges now repel and the repulsion grows stronger as the domain wall approaches the node. Under the assumption of overdamped dynamics, the wall stops when the Coulomb repulsion between the charges reaches the level of the Zeeman force from the external field.  One might think that this may be an equilibrium situation, but we show as follows that this is not the case. The arriving domain wall generates a strong field at the node, whose magnitude is easy to estimate. Since the domain wall is in equilibrium, the force applied to it by the external field, $F = \mu_0 |Q_\mathrm{w}| H_\mathrm{c}$, is balanced by the Coulomb repulsion of the node. By Newton's third law, the domain wall applies an equal force to the node.  The field created by the wall at the node is $H = F/|\mu_0 Q_j| = |Q_\mathrm{w}/Q_j| H_\mathrm{c} = 2H_\mathrm{c}$. This field is added to the externally applied field $H_c$. The resulting field is sufficiently strong to trigger the emission of another domain wall from the node. (This works for any relevant direction of the applied field.) The charge of node $j$ changes sign, $q_j = \mp 1 = -q_\mathrm{w}/2$, and subsequently absorbs the stopped domain wall. 

\subsection{Basic physics: quenched disorder}
\label{sec:basics-disorder}

Imperfections of magnetic links and junctions create local variations of the critical field $H_\mathrm{c}$. If the variations of $H_\mathrm{c}$ result from a large number of small errors, one expects a Gaussian distribution of critical fields $\rho(H_c)$ with a mean $\bar H_\mathrm{c}$ and a width $\delta H_\mathrm{c}$ given by
\begin{equation} 
\rho(H_\mathrm{c}) 
	= \frac{1}{\sqrt{2\pi}\delta H_\mathrm{c}} 
		\exp{\left(-
				\frac{(H_\mathrm{c} - \bar H_\mathrm{c})^2}{2 \delta H_\mathrm{c}^2}
			\right)}.
\label{eq:Gaussian}
\end{equation}
In the limit of strong disorder, when the distribution width $\delta H_\mathrm{c}$ is comparable to the average $\bar H_\mathrm{c}$, nodes with the highest critical fields may fail to follow the scenario shown in Figure \ref{fig:reversal-link}(b) and remain in a state with a triple charge until the field becomes strong enough.  Nodes with triple charges have been observed by Ladak \textit{et al.} \cite{Ladak:2010, NJP.13.023023}. In contrast, other samples have never shown triply charged defects \cite{qi:094418}, indicating that these samples are in the low-disorder limit, $\delta H_\mathrm{c} \approx 0.04 \bar H_\mathrm{c}$ \cite{PhysRevLett.107.167201}. 

The distribution width $\delta H_\mathrm{c}$ can be compared to another characteristic field strength, the magnetic field generated by an adjacent node, $H_0 = Mtw/(4\pi l^2)$. With the aid of Equation (\ref{eq:Hc-estimate}), we estimate
\begin{equation}
H_0/H_\mathrm{c} = (a/l)^2 \approx (w/l)^2.
\label{eq:H0-estimate}
\end{equation}
If $H_0 \ll \delta H_\mathrm{c}$, the Coulomb fields produced by adjacent and more distant nodes can be ignored to a first approximation. The Coulomb contribution to the net field on a given site is small, but occasionally the redistribution of magnetic charges on nearby sites may trigger the emission of a domain wall if the net field is close to the critical value. See Sec.~\ref{sec:numerics-120} for further details. In the opposite limit, $H_0 \gg \delta H_\mathrm{c}$, these internal fields must be taken into account. The reversal of magnetization on one link alters the magnetic charges on its ends. The resulting increments of the total magnetic field at nearby nodes, of order $H_0$, may be sufficient to trigger the emission of domain walls from them. Samples we studied previously \cite{qi:094418, PhysRevLett.107.167201} appear to be in the regime where $H_0$ and $\delta H_\mathrm{c}$ are comparable.

\section{Microscopic basis for the model}
\label{sec:micro}

To test the basic model of magnetization dynamics presented in Sec.~\ref{sec:basics}, we performed numerical simulations of magnetization dynamics in a small portion of the honeycomb network by using the micromagnetic simulator OOMMF \cite{oommf}. 

\begin{figure}
\begin{center}
\includegraphics[width=0.32\columnwidth]{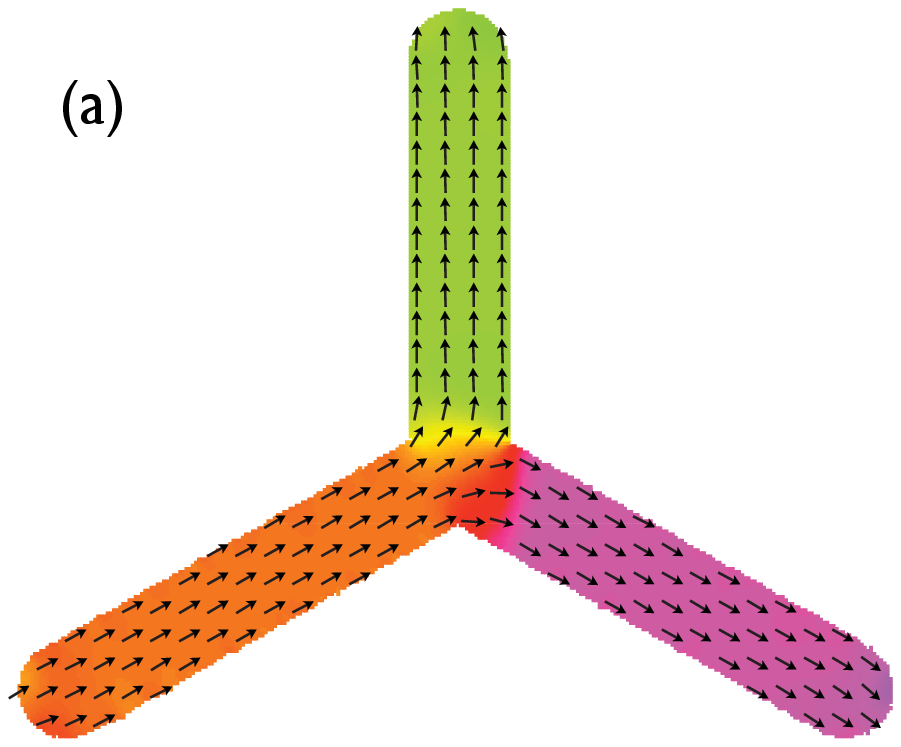}
\includegraphics[width=0.32\columnwidth]{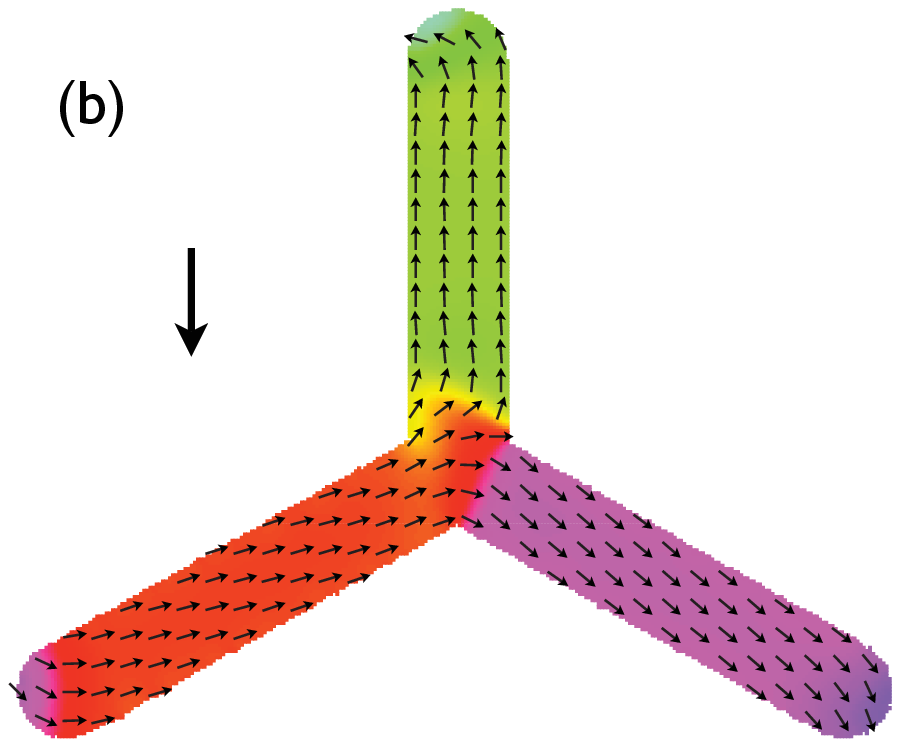}
\includegraphics[width=0.32\columnwidth]{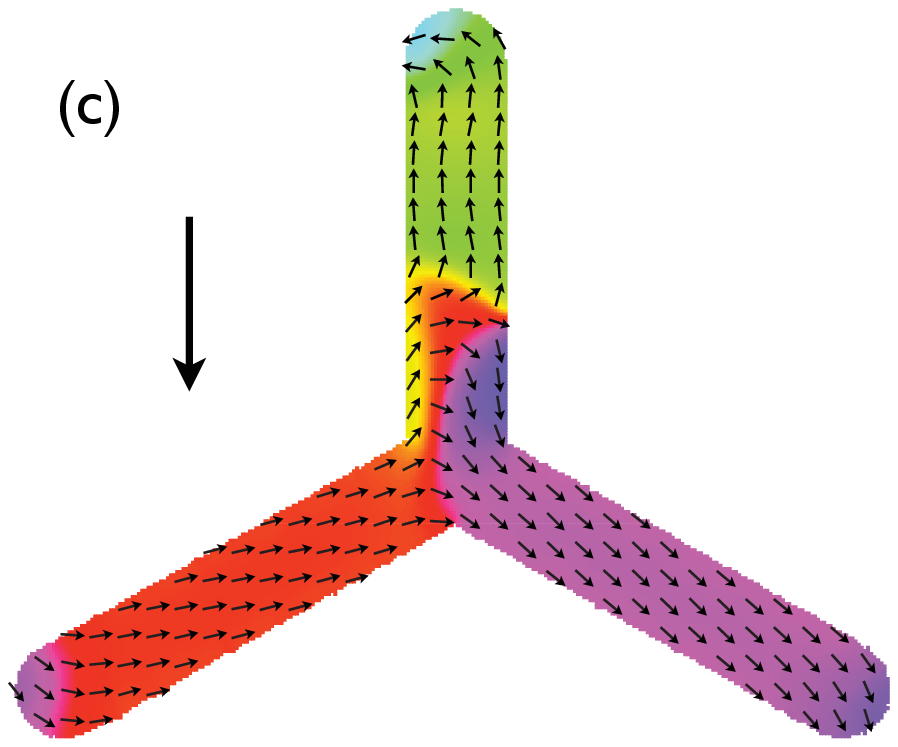}
\includegraphics[width=0.32\columnwidth]{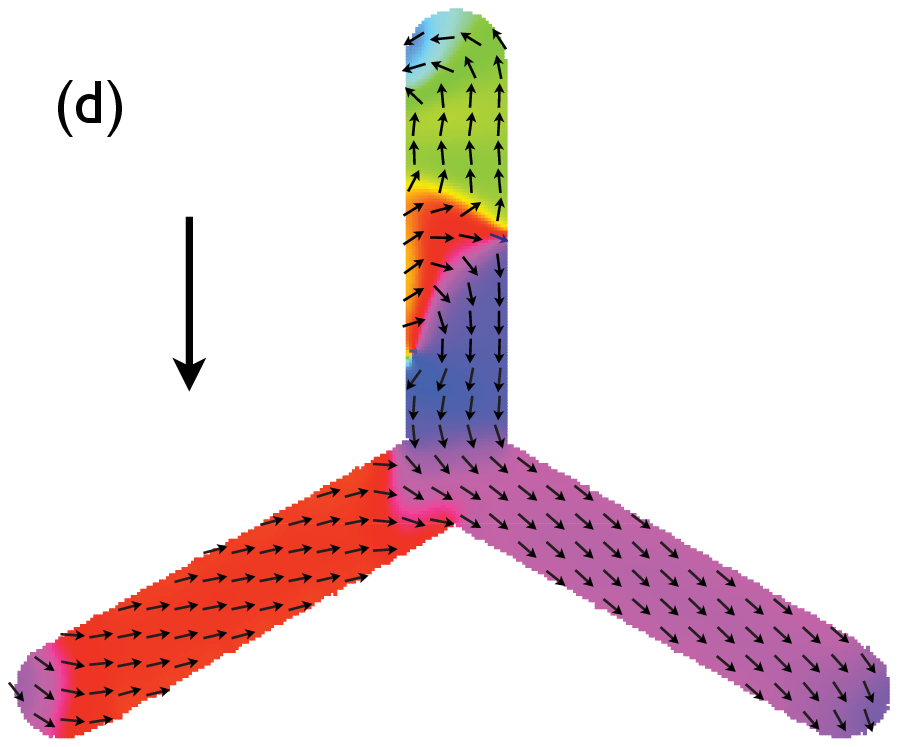}
\includegraphics[width=0.32\columnwidth]{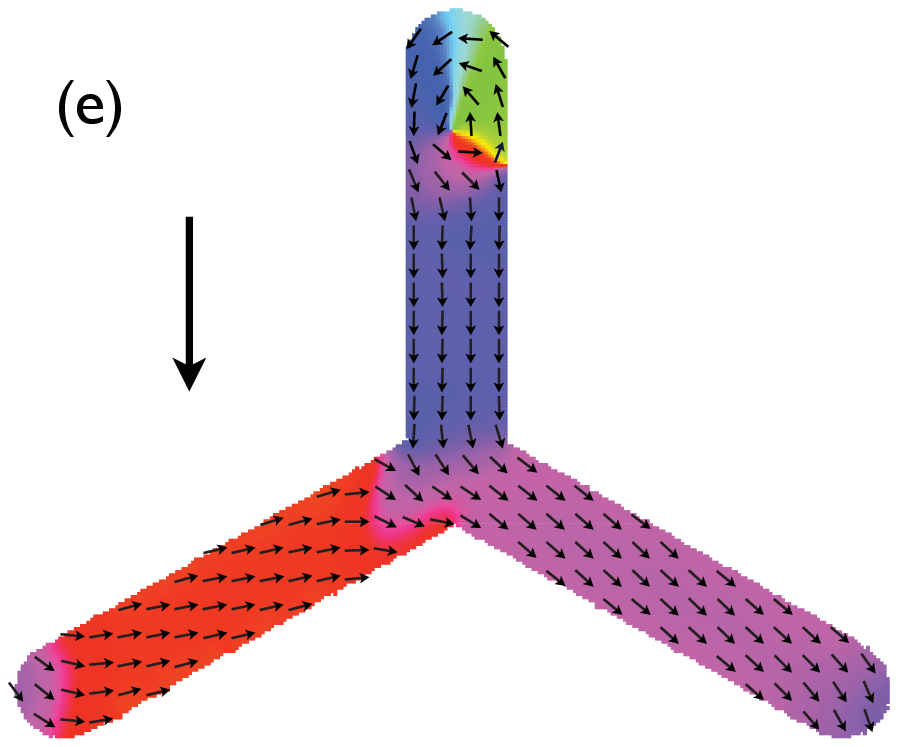}
\includegraphics[width=0.32\columnwidth]{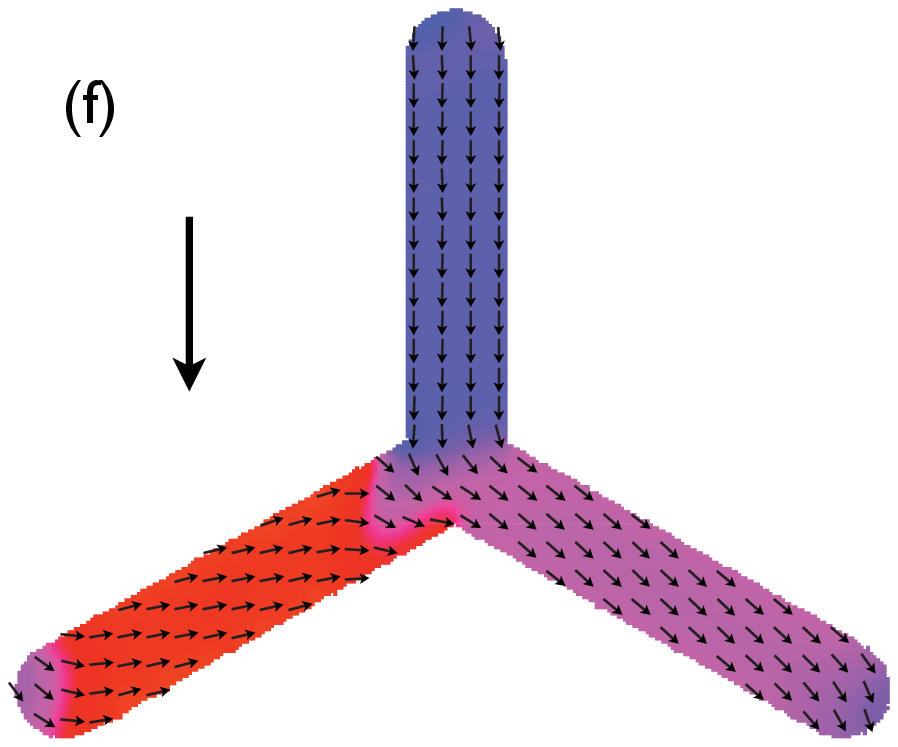}
\caption{Reversal of magnetization in a magnet consisting of three joint links in an applied magnetic field (vertical arrow). In panels (a) through (c), the strength of the field slowly increases from 0 to a critical value as the magnetization adjusts adiabatically. In panels (c) through (f), a domain wall detaches from the node and quickly propagates through the vertical link; the field value remains essentially unchanged. Micromagnetic simulation (oommf).
} 
\label{fig:reversal-micro}
\end{center}
\end{figure}

\begin{figure}
\begin{center}
\includegraphics[width=0.7\columnwidth]{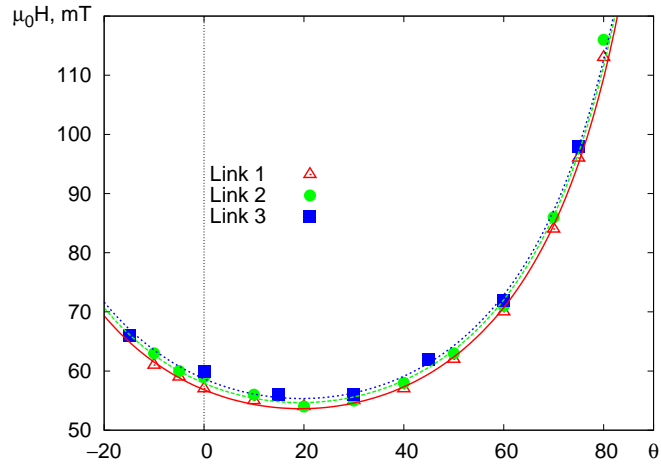}
\caption{The dependence of the critical field $H$ on the angle $\theta$ between the field and the link. The lines are best fits to Equation (\ref{eq:Hc-theta}).  Links 1, 2, and 3 had $H_{c} = 53.6$ mT, 54.7 mT, and 55.3 mT and $\alpha = 19.3^\circ$, $19.4^\circ$, and $19.4^\circ$, respectively. The same numerical experiments were repeated three times, with the field initially lined up with Link 1, 2, or 3, and then rotated through $180^\circ + \theta$ from that direction.}
\label{fig:H-vs-theta}
\end{center}
\end{figure}

The typical numerical experiment involved a junction of three permalloy magnetic wires of length $l = 500$ nm, width $w = 110$ nm, and thickness $t= 23$ nm \cite{qi:094418}.  We used the two-dimensional version of the \texttt{oommf} code with cells 2 nm $\times$ 2 nm $\times$ 23 nm. (The lateral size of the unit cell should not exceed the minimal length in the micromagnetic problem, the magnetic exchange length obtained from exchange and dipolar couplings. In permalloy, it is about 5 nm \cite{ThiavilleBook06}.) The magnetization field $\mathbf M(\mathbf r)$ was allowed to relax to an equilibrium state with magnetic charge $q = \pm1$ at the junction, Figure \ref{fig:reversal-micro}. An external magnetic field was then applied in a fixed direction and its magnitude was slowly increased keeping the system in a state of local equilibrium. Eventually, the magnet reached a point of instability when a domain wall was emitted from either the central junction or one of the peripheral ends of the wires, depending on the direction of the applied field. The wall then propagated to the opposite end of the link reversing the link's magnetization.  Using those orientations of the field for which a domain wall is emitted from the junction, we determined the dependence of the critical field $H$ on the angle $\theta$ between the field and the link in which the reversal occurs, Figure \ref{fig:H-vs-theta}.  

Two features of the angular dependence in Figure \ref{fig:H-vs-theta} stand out. First, $H(\theta)$ is not an even function of the angle $\theta$, and contrary to our expectations, the critical field is not at its lowest when the field is parallel to the link. Second, the critical fields for three different links in the experiment have the same shape but differ in the overall scale $H_\mathrm{c}$. 

We have traced the physical origin for the asymmetric dependence of the critical field $H(\theta)$ to an asymmetric distribution of magnetization at the junction, Figure \ref{fig:reversal-micro}. The energetics of the emission process shown in the figure can be described in the language of collective coordinates \cite{clarke:134412}. The soft mode associated with the emission of a domain wall into the vertical link is the domain wall displacement $X$ along the link.  To the first order in the applied field $\mathbf H$ and to the second order in $X$, the energy is
\begin{equation}
U(H,X) = U(0,0) - \mu_0 X (Q_{xx}H_x + Q_{xy} H_y) + kX^2/2, 
\end{equation}
where $Q_{xx}$, $Q_{xy}$ and $k$ are phenomenological constants. Generally speaking, the off-diagonal component $Q_{xy}$ does not vanish unless the magnetization distribution is symmetric under the reflection $y \mapsto -y$. The equilibrium position of the wall depends on the direction of the applied field $\mathbf H = (H \cos{\theta}, H\sin{\theta}, 0)$ as follows: 
\begin{equation}
X_\mathrm{eq} = (\mu_0/k)(Q_{xx}H_x + Q_{xy} H_y) = (\mu_0/k) \tilde Q H \cos{(\theta-\alpha)},
\label{eq:X-eq}
\end{equation}
where the offset angle $\alpha$ and effective charge $\tilde Q$ are defined through 
\begin{equation}
Q_{xx} = \tilde Q \cos{\alpha}, \quad 
Q_{xy} = \tilde Q \sin{\alpha}.
\end{equation}
According to Equation (\ref{eq:X-eq}), the relevant component of the magnetic field $\mathbf H$ is found by projecting the field onto the \textit{easy axis} of a (majority) link, which is rotated through angle $\alpha$ toward the minority link. These considerations suggest the following modification for the postulated field dependence of the critical field (\ref{eq:Hc-theta-estimate}):
\begin{equation}
H_\mathrm{c}(\theta) = H_\mathrm{c}/\cos{(\theta - \alpha)}.
\label{eq:Hc-theta}
\end{equation}
As Figure \ref{fig:H-vs-theta} shows, this equation provides a good description of the angular dependence of the critical field with the offset angle $\alpha \approx 19^\circ$. The overall scale of the critical field $H_\mathrm{c}$ showed variations reflecting small imperfections of links in the simulation. For instance, the square lattice of magnetic moments used in oommf simulations is incommensurate with links pointing at $60^\circ$ to a lattice axis and creates edge roughness.  This observation confirms the proposed model of disorder introduced in Sec.~\ref{sec:basics-disorder}. 

\section{Numerical simulations}
\label{sec:numerics}

\begin{figure}
\begin{center}
\includegraphics[width=0.45\columnwidth]{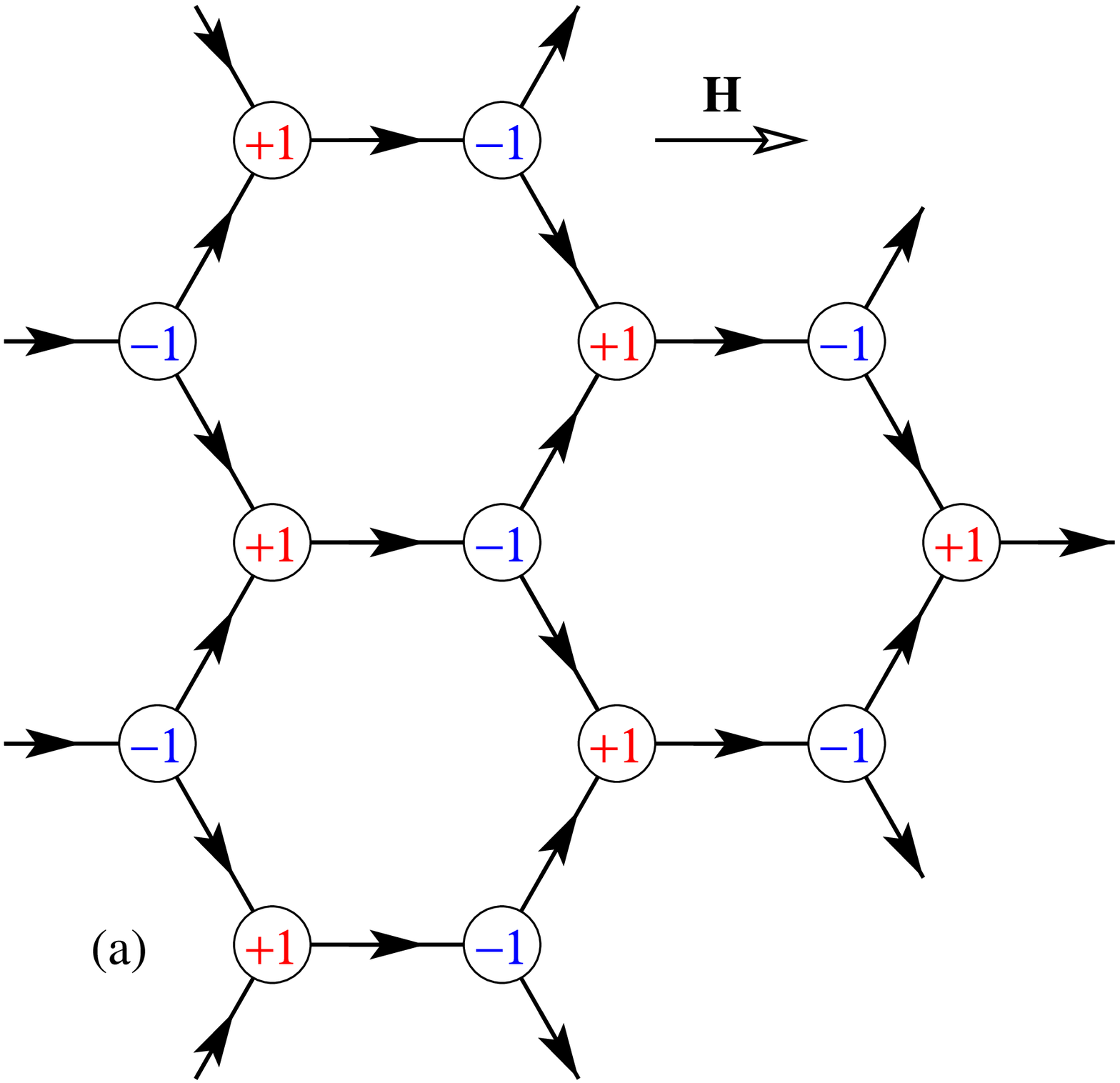}
\includegraphics[width=0.45\columnwidth]{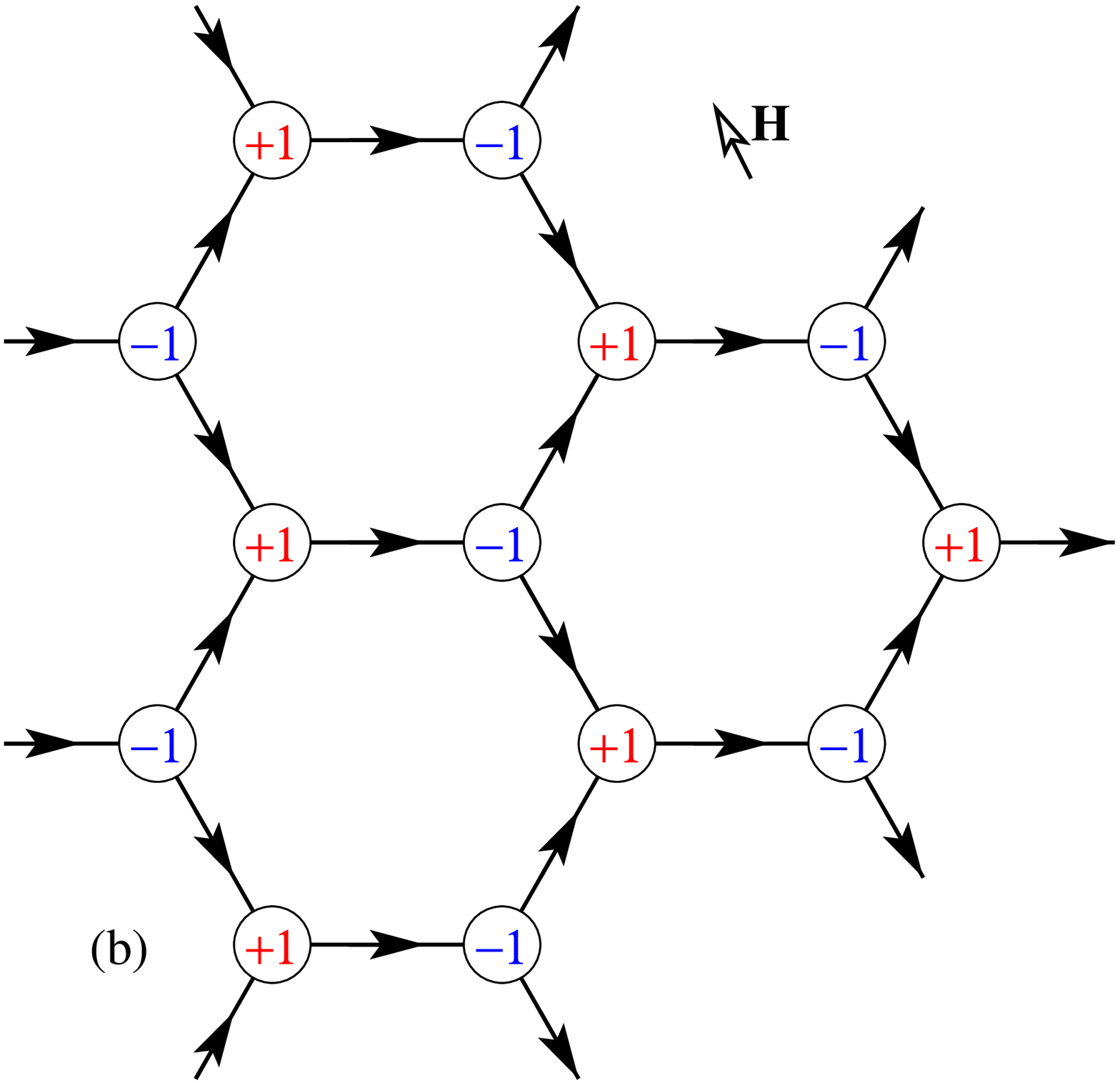}

\bigskip
\includegraphics[width=0.45\columnwidth]{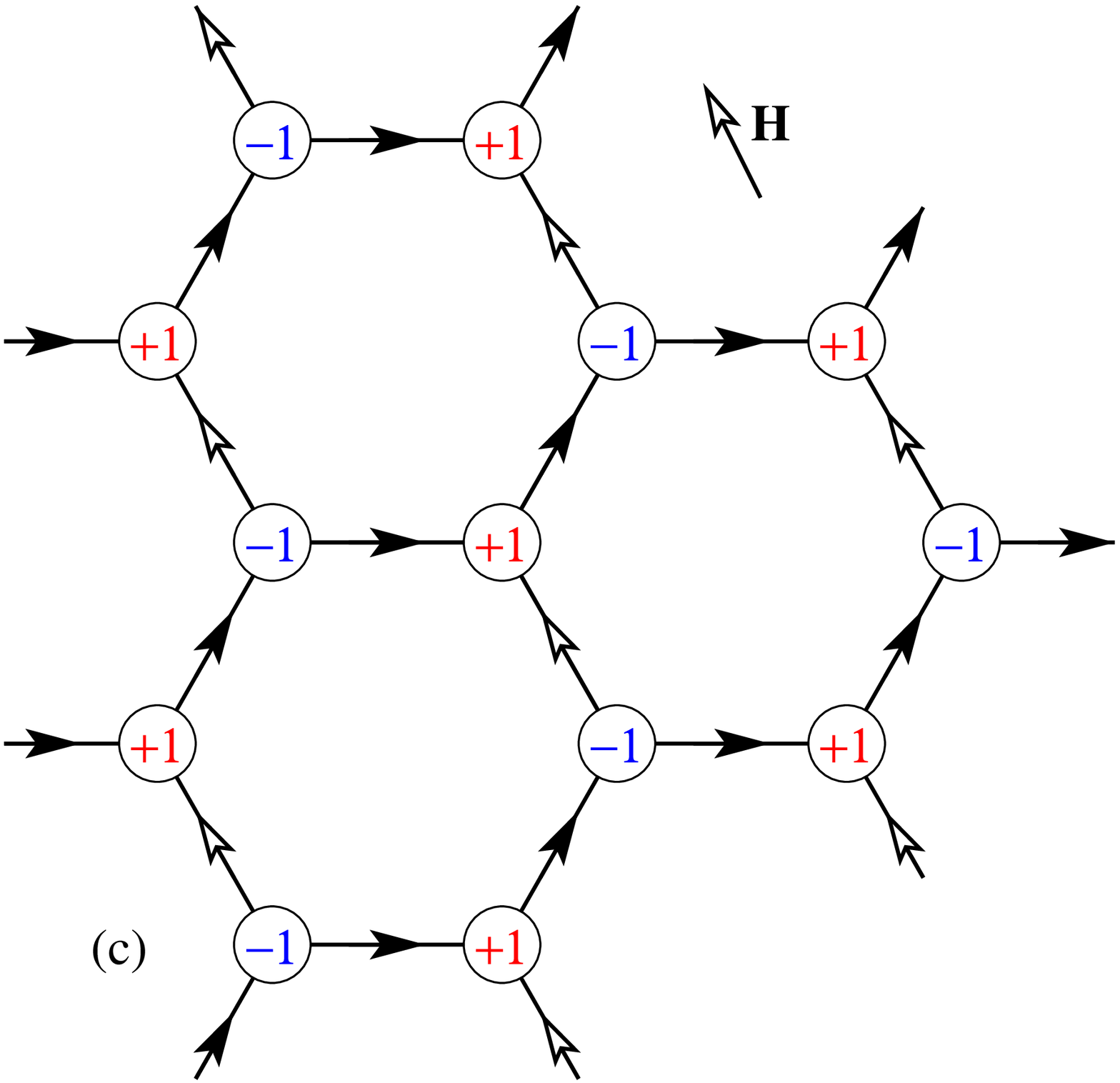}
\includegraphics[width=0.45\columnwidth]{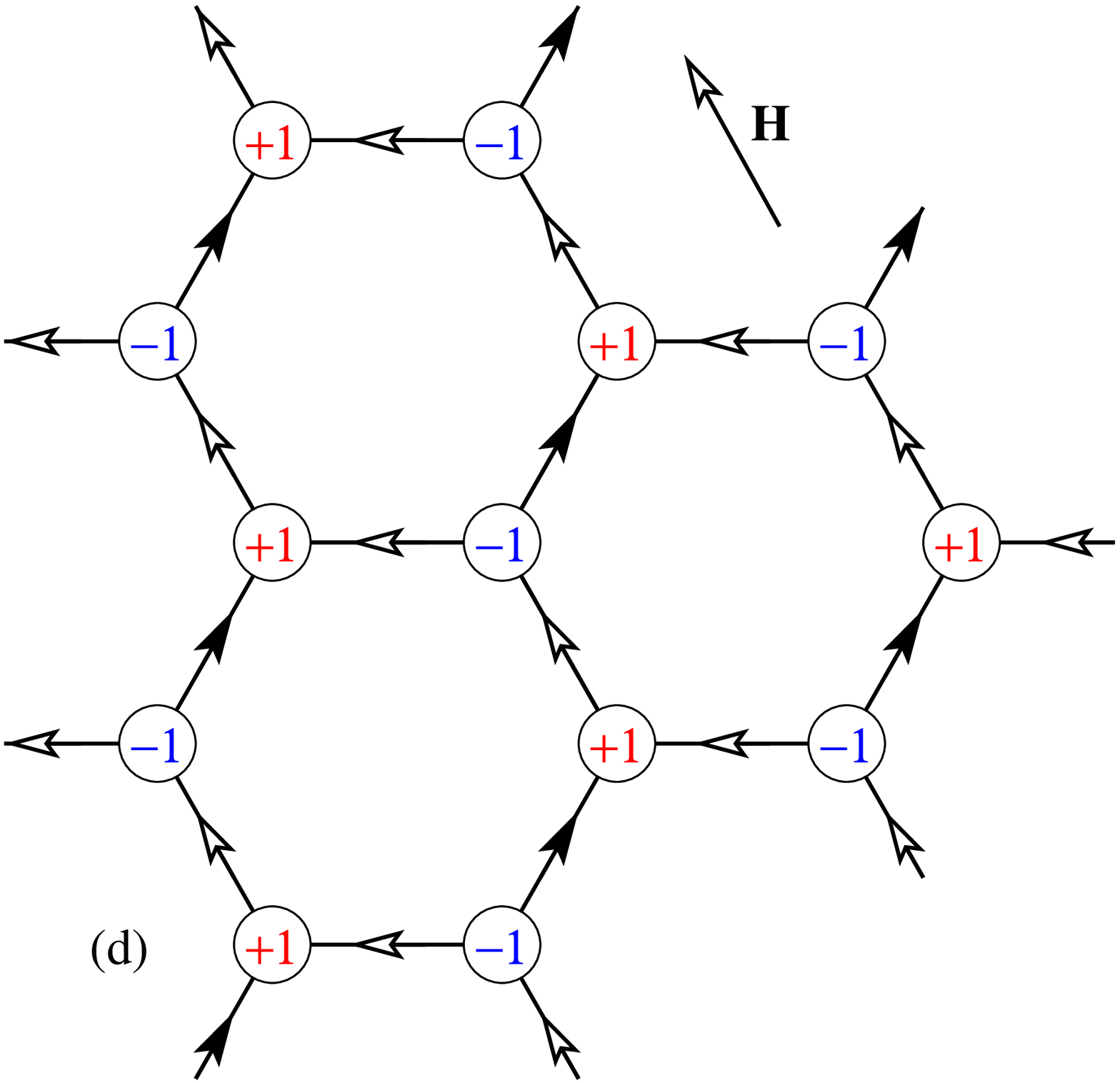}
\end{center}
\caption{Magnetization reversal in an applied magnetic field. (a) The system is initially magnetized in a strong horizontal field. (b-d) The field is then switched off and applied at $120^\circ$ to the original direction with a gradually increasing magnitude. Open arrows denote links with reversed magnetization. }
\label{fig:reversal-120}
\end{figure}

The heuristic considerations of Section~\ref{sec:basics} and the micromagnetic simulations of Section~\ref{sec:micro} suggested a coarse-grained model of magnetization dynamics in which the basic degrees of freedom are Ising variables of magnetization $\sigma_{ij}$ on links and magnetic charges $q_i$ on sites of the honeycomb lattice. Each link has its own fixed critical field $H_\mathrm{c}$. The critical fields form a Gaussian distribution (\ref{eq:Gaussian}) of width $\delta H_\mathrm{c}$ around the mean $\bar H_\mathrm{c}$.  The average, $\bar H_\mathrm{c} = 50$ mT, was chosen on the basis of our micromagnetic simulations, whereas the relative width was set to $\delta H_\mathrm{c}/\bar H_\mathrm{c} = 0.05$, a value inspired by our experimental observations \cite{PhysRevLett.107.167201}. Simulations were performed in a rectangular sample with 937 links. The edge consisted of ``dangling'' links with no other links attached to their external ends. We choose the initial state with a maximum total magnetization that can be obtained by placing the system in a strong magnetic field along one set of links, Figure~\ref{fig:reversal-120}(a). Simulation details are described in \ref{sec:app-procedure}.

Following initialization, the external field is switched off and reapplied along a different direction, at an angle $\theta$ to its initial orientation, Figure~\ref{fig:reversal-120}(b-d). To stimulate magnetization dynamics, the rotation angle must be large enough so that $\mathbf H$ would have a negative projection onto at least some of the majority links. When $|\theta|$ is between $30^\circ$ and $90^\circ$, only one of the three sublattices of links will reverse. Two sublattices reverse when $|\theta|$ is between $90^\circ$ and $150^\circ$. The entire lattice undergoes a reversal when $|\theta| > 150^\circ$. 

\begin{figure}
\begin{center}
\includegraphics[width=0.3\columnwidth]{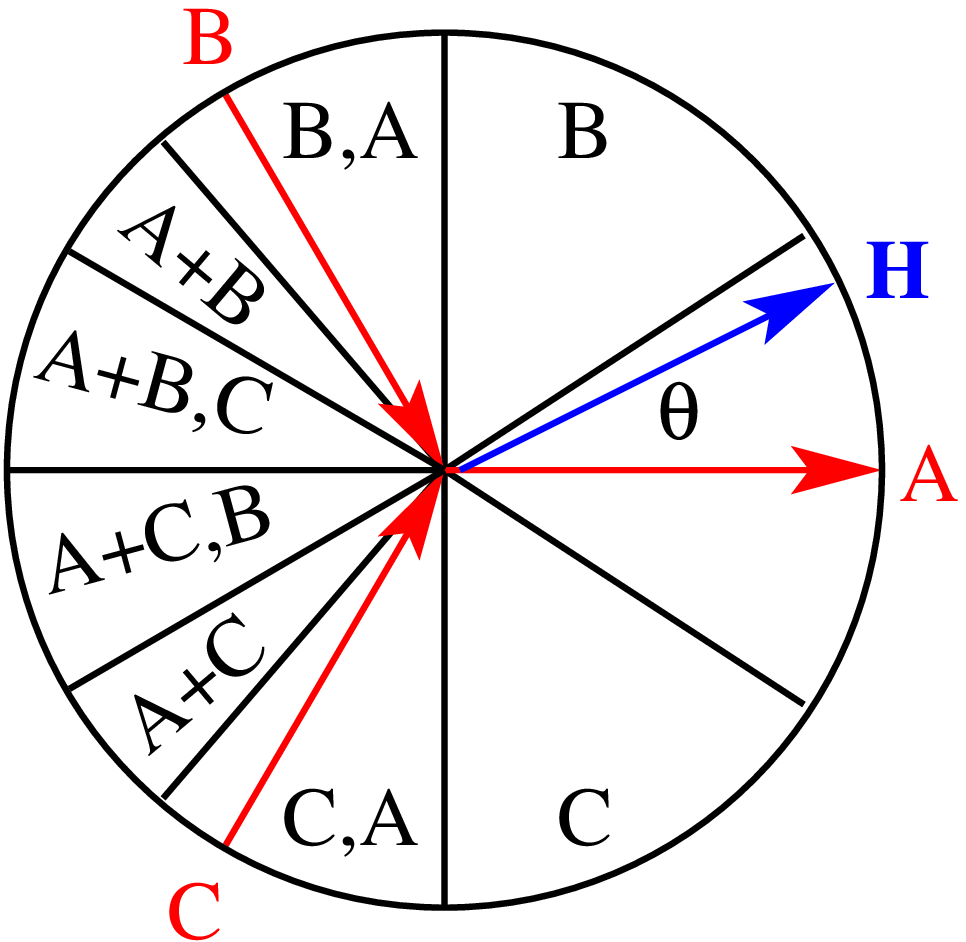}
\includegraphics[width=0.49\columnwidth]{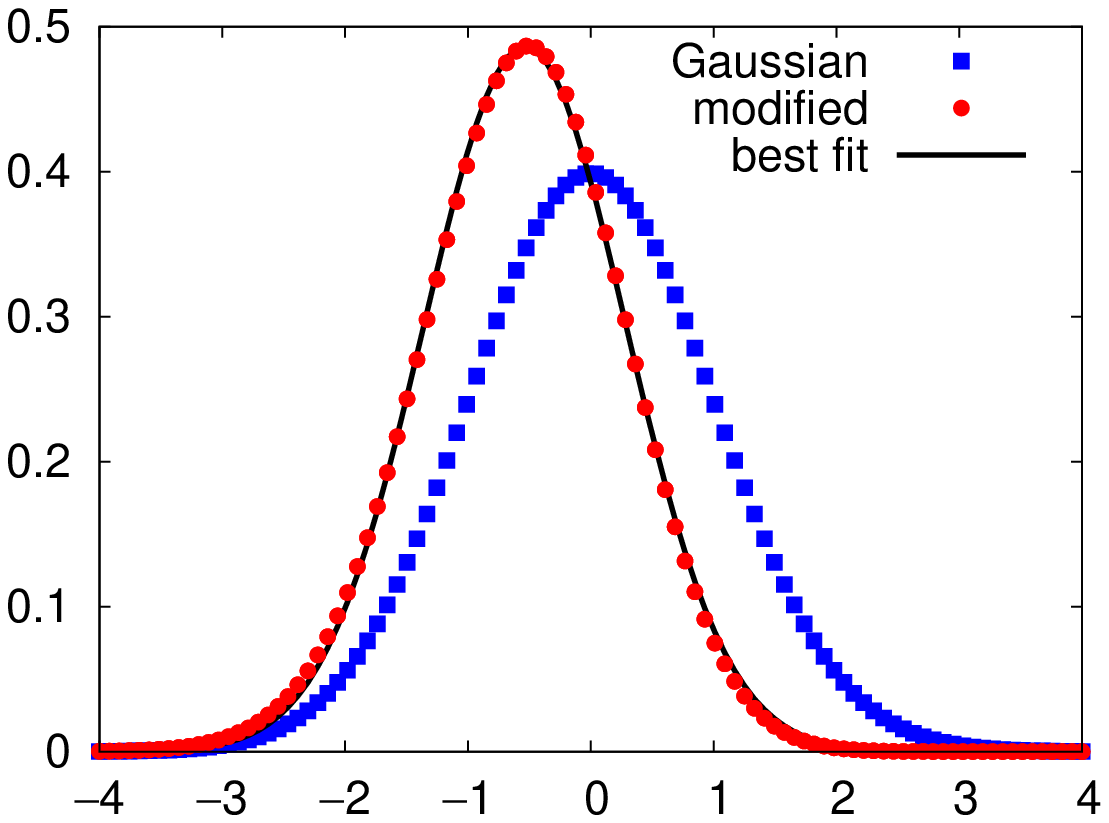}
\end{center}
\caption{Left panel: Regimes of magnetization reversal. $0 < \theta < 30^\circ$: no reversal. $30^\circ < \theta < 90^\circ$: sublattice B only. $90^\circ < \theta < 131^\circ$: B, then A. $131^\circ < \theta < 150^\circ$: A and B reverse together. $150^\circ < \theta < 180^\circ$: A and B, then C. Similar regimes obtain for negative $\theta$, with sublattices B and C exchanged. Right panel: Illustration to Equation (\ref{eq:Gaussian-modified}).  The Gaussian distribution $\exp{(-x^2/2)}/\sqrt{2\pi}$ (blue squares), the modified distribution $\mathrm{erfc}(x/\sqrt{2}) \, \exp{(-x^2/2)}/\sqrt{2\pi}$ (red circles), and the best Gaussian approximation, $\exp{(-(x-\delta)^2/2\sigma^2)}/\sqrt{2\pi}\sigma$ with the mean $\delta = -0.54$ and width $\sigma = 0.82$ (solid line).}
\label{fig:regimes}
\end{figure}

Aside from the number of active sublattices, there are marked differences in the dynamics of the reversal. For small angles of rotation, $|\theta| < 131^\circ$, the reversals occur in a gradual and uncorrelated manner, with individual links switching when the applied field reaches the link's critical field. For larger angles, $|\theta| > 131^\circ$, we observed \textit{avalanches} in which long chains of links reverse magnetization simultaneously. This kind of switching happens when the sublattice whose magnetization is most antiparallel to the applied field cannot switch first because it consists entirely of minority links and must wait for one of the other sublattices to begin its reversal. If that happens in a higher field, the former sublattice acts like a loaded spring, making the reversal nearly instantaneous. 
A diagram depicting different regimes as a function of the field rotation angle $\theta$ is shown in Figure~\ref{fig:regimes}. 

In the simplest case, the reversal of magnetization in a link occurs when the magnetic field reaches the critical value for that link. The links thus reverse on an individual basis, largely independently from the others (but see below). To be more precise, the two ends of a link have different critical fields and the reversal begins from the end with the lower critical field and stops at the other end. The effective probability density of the critical fields thus changes from a Gaussian distribution to 
\begin{eqnarray}
f'(H_\mathrm{c}) &=& 2\rho(H_\mathrm{c}) \int_{H_\mathrm{c}}^\infty \mathrm{d}H \, \rho(H)
\nonumber\\
&=& \frac{1}{\sqrt{2\pi}\delta H_\mathrm{c}} 
		\exp{\left(-
				\frac{(H_\mathrm{c} - \bar H_\mathrm{c})^2}{2 \delta H_\mathrm{c}^2}
			\right)} \, \mathrm{erf}\left(\frac{H_\mathrm{c}-\bar H_\mathrm{c}}{\delta H_\mathrm{c} \sqrt{2}}\right).
\label{eq:Gaussian-modified}
\end{eqnarray}
It can be seen in Figure \ref{fig:regimes} (right panel) that the resulting distribution is very close to a Gaussian with renormalized mean and width,
\begin{equation}
\bar H_\mathrm{c}' = \bar H_\mathrm{c} - 0.54 \delta H_\mathrm{c}, 
\quad
\delta H'_c = 0.82 \delta H_\mathrm{c}.
\label{eq:Gaussian-renorm}
\end{equation}
In our simulation, the renormalized values are $\bar H'_c = 48.7$ mT and $\delta H'_c/\bar H'_c = 0.042$.

\subsection{$30^\circ < \theta < 131^\circ$: gradual reversal}
\label{sec:numerics-120}

With the field rotated through $\theta = 120^\circ$, two sets of links have a negative projection of magnetization onto the field. In Figure~\ref{fig:reversal-120}(b), they are the horizontal minority links and the majority links parallel to the field.  Because emission of a domain wall into a minority link requires a very high field, it is the majority links that undergo magnetization reversal first.  The field makes an angle $\alpha \approx 19^\circ$ with their easy axes, so the reversal is expected to occur around the field $H_1 = \bar H'_c/\cos{(-19^\circ)} = 51.5$ mT. 

Magnetization reversal in the links parallel to the field alters the magnetic charges on all sites, Figure~\ref{fig:reversal-120}(c). As a result of this change, horizontal links join the majority and become capable of reversing their magnetization. The external field makes an angle $60^\circ - \alpha \approx 41^\circ$ with their easy axes, so their magnetization reversal is expected to occur when the field reaches a higher value, $H_2 = \bar H'_c/\cos{41^\circ} = 64.5$ mT. In the presence of disorder, the reversal regions are expected to have finite widths, $\delta H_1/H_1 = \delta H_2/H_2 = \delta H'_\mathrm{c}/\bar H'_\mathrm{c}$. For a Gaussian distribution of critical fields, magnetization measured along the applied field is expected to be a superposition of error functions:
\begin{equation}
\frac{M(H)}{M_\mathrm{max}} 
	= \frac{1}{2}\mathrm{erf}\left(\frac{H-H_1}{\delta H_1 \sqrt{2}}\right)
	+ \frac{1}{4}\mathrm{erf}\left(\frac{H-H_2}{\delta H_2 \sqrt{2}}\right)
	+ \frac{1}{4}.
\label{eq:M-vs-H-120}
\end{equation}
The three terms reflect the contributions of the three groups of links with different orientations.

\begin{figure}
\begin{center}
\includegraphics[width=0.49\columnwidth]{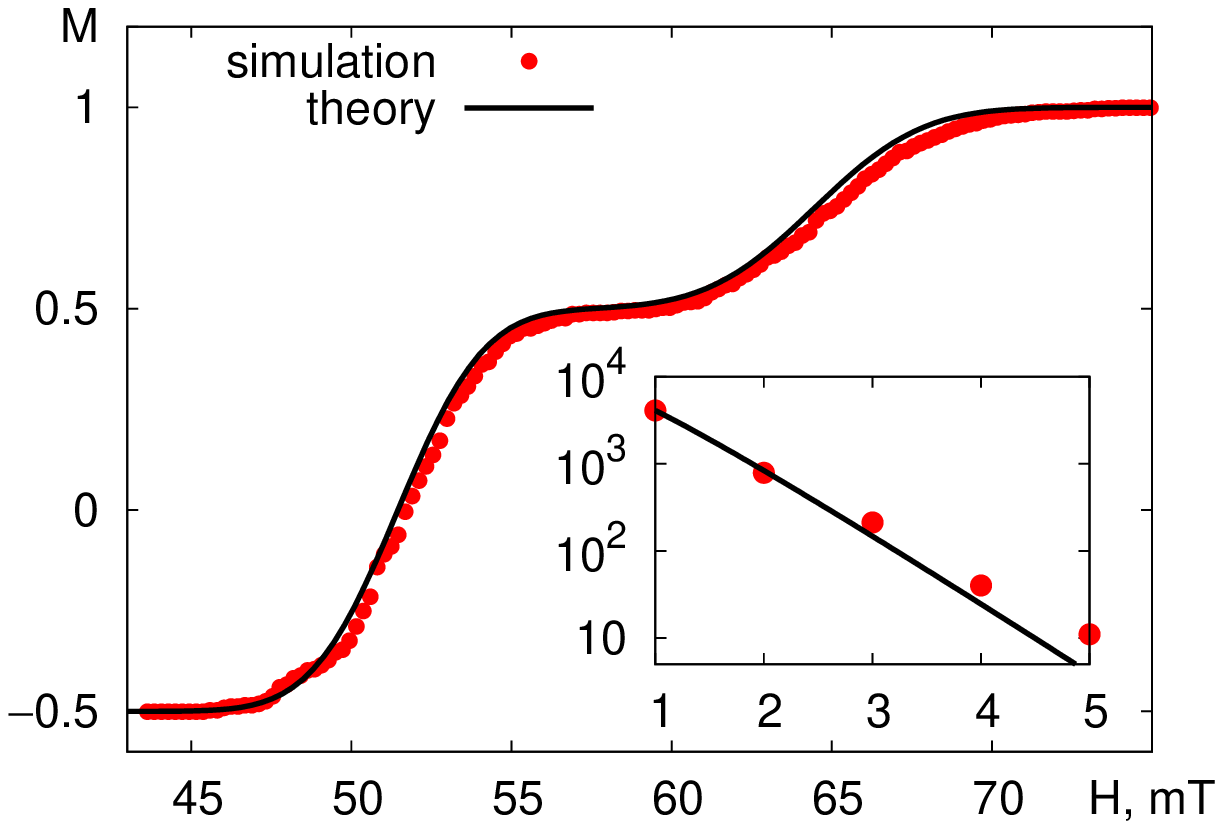}
\includegraphics[width=0.49\columnwidth]{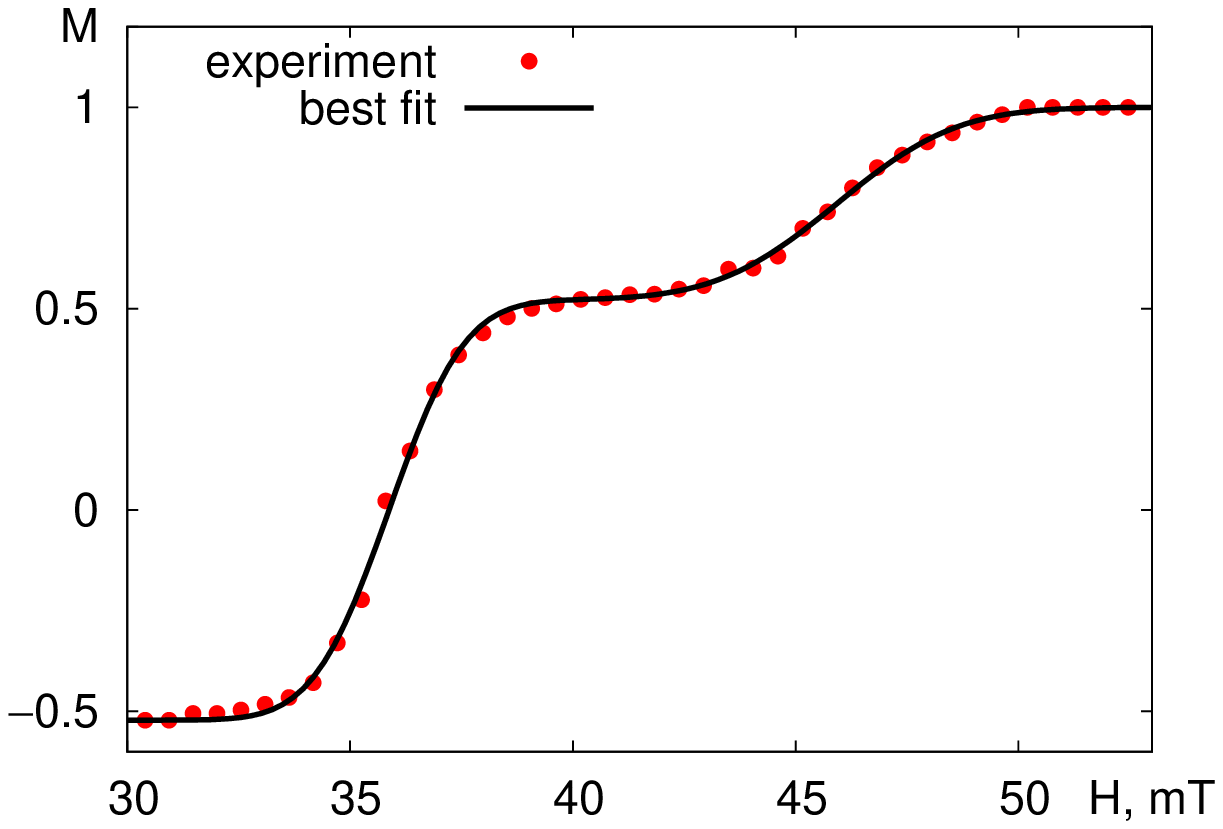}
\end{center}
\caption{Magnetization reversal curve $M(H)$ in an applied field rotated through $120^\circ$. Left: Simulated magnetization curve $M(H)$ (red circles) is well approximated by the theoretical curve (\ref{eq:M-vs-H-120})  (solid black line).  Inset: semi-log plot of the number of avalanches as a function of their length. Right: Experimental magnetization curve $M(H)$ (red circles) \cite{PhysRevLett.107.167201} and the best fit to Eq.~(\ref{eq:M-vs-H-120}) (solid black line). }
\label{fig:simulations-120}
\end{figure}

The simulated dependence $M(H)$ is shown in Figure \ref{fig:simulations-120} along with the theoretical curve (\ref{eq:M-vs-H-120}) that takes into account the renormalization of the Gaussian distribution parameters (\ref{eq:Gaussian-renorm}).

A close inspection of the simulated curve $M(H)$ shows that on occasion several adjacent links reverse simultaneously due to a positive feedback during the reversal. When magnetization of a link is reversed, magnetic charges at its ends are switched. The net magnetic field on an adjacent site, projected onto its easy axis, increases by 
\begin{equation}
\Delta H = 2 H_0 \cos{41^\circ} - (2H_0/3) \cos{11^\circ} = 0.86 H_0 = 0.74 \mbox{ mT.}
\end{equation} 
The extra field is not negligible on the scale of the critical-field distribution width $\delta H'_c = 2.0$ mT. It can help to stimulate the emission of a domain wall at an adjacent site if that site's critical field is not too high. This kind of positive feedback causes \textit{avalanches}, in which magnetization reversals occur nearly simultaneously in links residing along a one-dimensional path determined by the orientations of easy axes.  For example, an avalanche occurring in the background of a fully magnetized state of Figure~\ref{fig:reversal-120}(b) would travel along the vertical direction. In the limit of small feedback, $\Delta H \ll \delta H'_c$, the distribution of avalanche lengths is exponential. Indeed, if the link starting an avalanche of length $n$ has a critical field $H$, $n-1$ of its neighbors must have critical fields in the range between $H$ and $H + \Delta H$.  The probability to find such a collection of links is
\begin{equation}
P_n \sim n \int [\rho(H) \Delta H]^{n-1} \rho(H) \, \mathrm{d}H 
= n^{1/2} \left( \frac{\Delta H}{\sqrt{2\pi} \delta H'_c} \right)^{n-1}
\label{eq:avalanche-stats-120}
\end{equation}
for a Gaussian distribution of critical fields (\ref{eq:Gaussian}). The distribution of avalanches seen in the simulation is shown in the inset of Figure~\ref{fig:simulations-120} along with the theoretical distribution (\ref{eq:avalanche-stats-120}).

These results can be directly compared to the experimental reversal curve measured in the same geometry \cite{PhysRevLett.107.167201}, Figure~\ref{fig:simulations-120} (right panel).  Although the overall scale of the magnetic field is substantially lower, the data are well fit by Eq.~(\ref{eq:M-vs-H-120}) with $H_1 = 35.9$ mT and $H_2 = 45.9$. The ratio of the reversal fields, $H_2/H_1 = 1.28$, agrees well with the theoretical value $H_2/H_1 = \cos{(-19^\circ)}/\cos{41^\circ} = 1.25$.  The relative widths are $\delta H_1/H_1 = 0.037$ and $\delta H_2/H_2 = 0.046$. 

The magnetization curve $M(H)$ was also measured experimentally \cite{PhysRevLett.107.167201} and simulated for $\theta = 100^\circ$, with similar results.  The experimentally measured reversal fields were $H_1 = 34.7$ mT and $H_2 = 91.5$ mT and relative widths $\delta H_1/H_1 = 0.033$ and $\delta H_2/H_2 = 0.047$. The reversal field ratio was $H_2/H_1 = 2.64$ in the experiment, somewhat off the theoretical value $H_2/H_1 = \cos{1^\circ}/\cos{61^\circ} = 2.06$.

Overall, it appears that our model provides a reasonably good description of magnetization reversal when the field is reapplied at $\theta = 120^\circ$ to the direction of initial magnetization. In this regime, the reversal proceeds in two well-defined stages, each involving one subset of links.  During each stage, links reverse largely independently, although sometimes the reversal in one link changes the field on a nearby site and triggers magnetization reversal there. The reversal fields are given approximately by the equations
\begin{equation}
H_1 = \bar H'_c/\cos{(120^\circ - \theta - \alpha)}, 
\quad
H_2 = \bar H'_c/\cos{(180^\circ - \theta - \alpha)}.
\label{eq:H1-H2}
\end{equation}
The reversal follows the two-stage scenario as long as $H_1 < H_2$, or $\theta < 150^\circ - \alpha = 131^\circ$. For larger field rotation angle $\theta$, the reversal proceeds in a very different manner. 

\subsection{$131^\circ < \theta < 180^\circ$: reversal with avalanches}
\label{sec:numerics-170}

\begin{figure}
\begin{center}
\includegraphics[width=0.45\columnwidth]{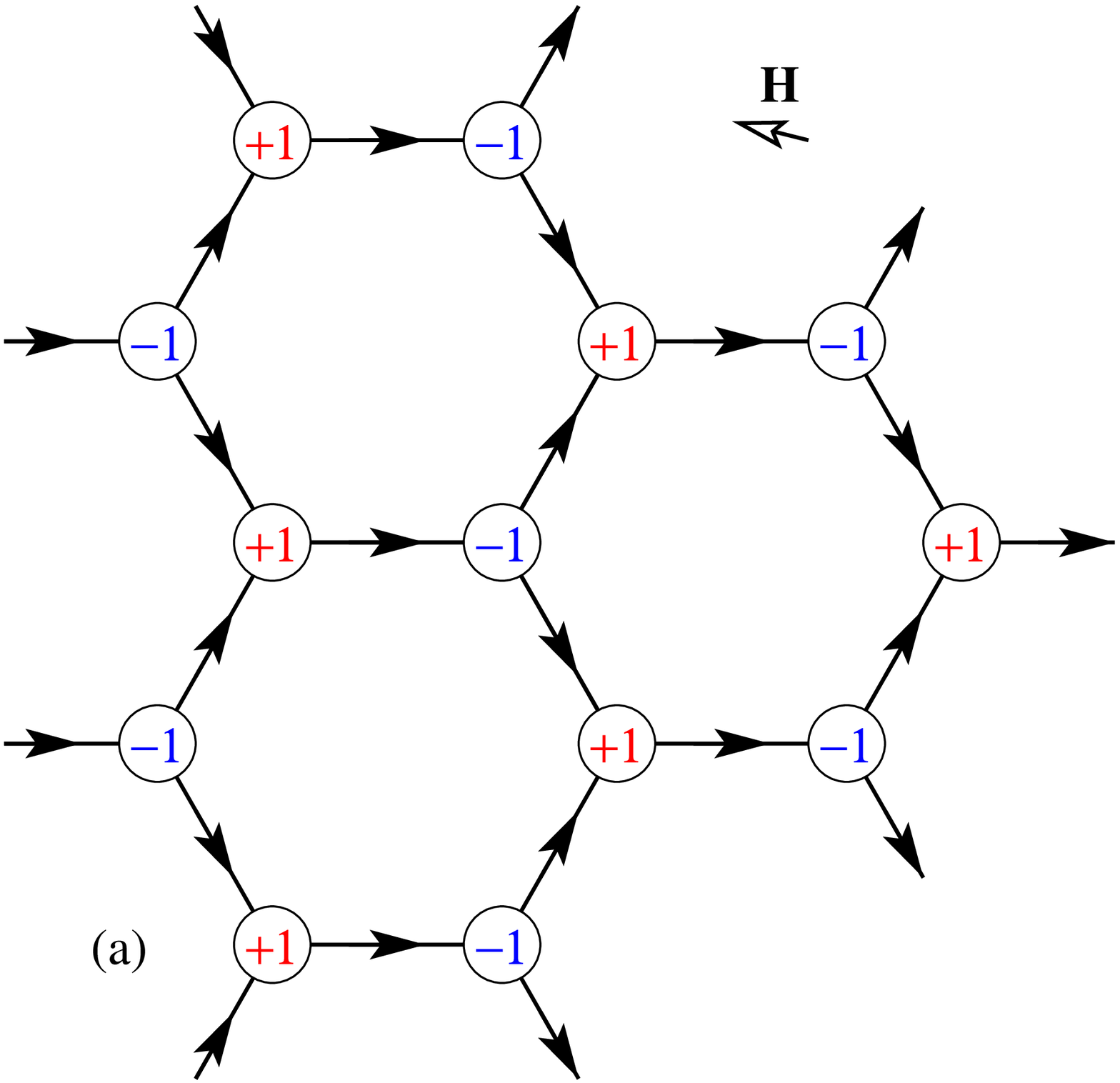}
\includegraphics[width=0.45\columnwidth]{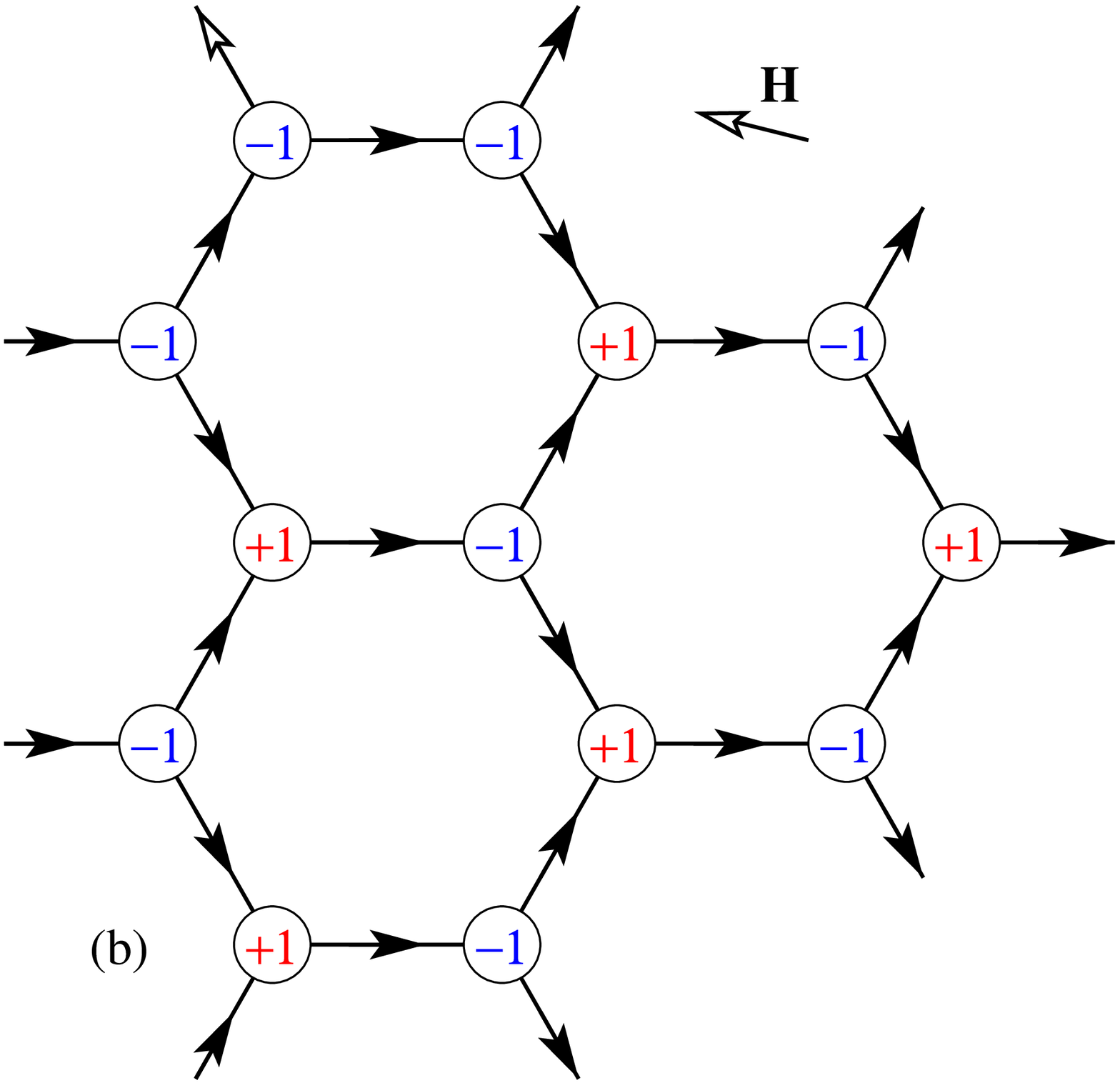}

\bigskip
\includegraphics[width=0.45\columnwidth]{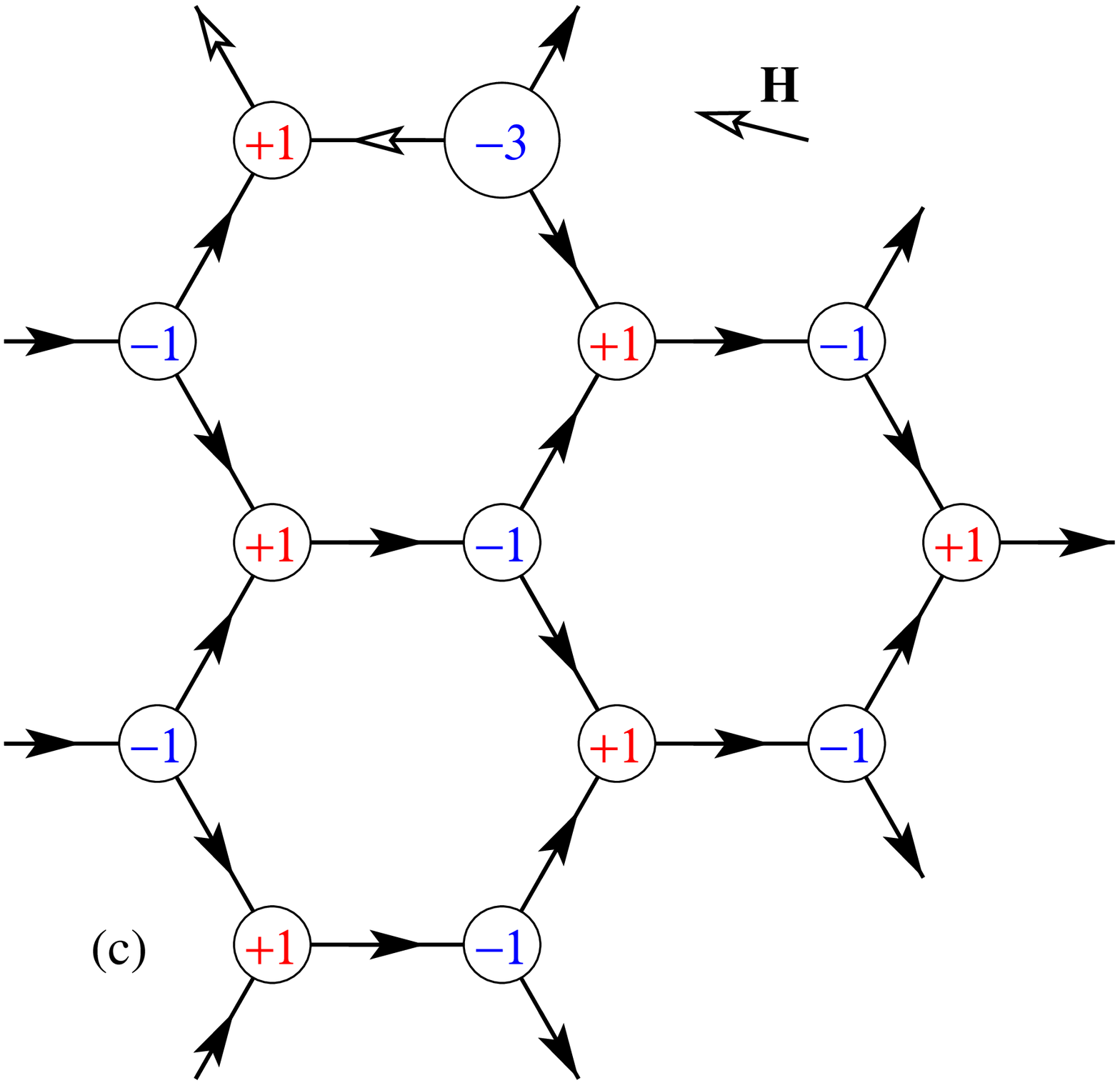}
\includegraphics[width=0.45\columnwidth]{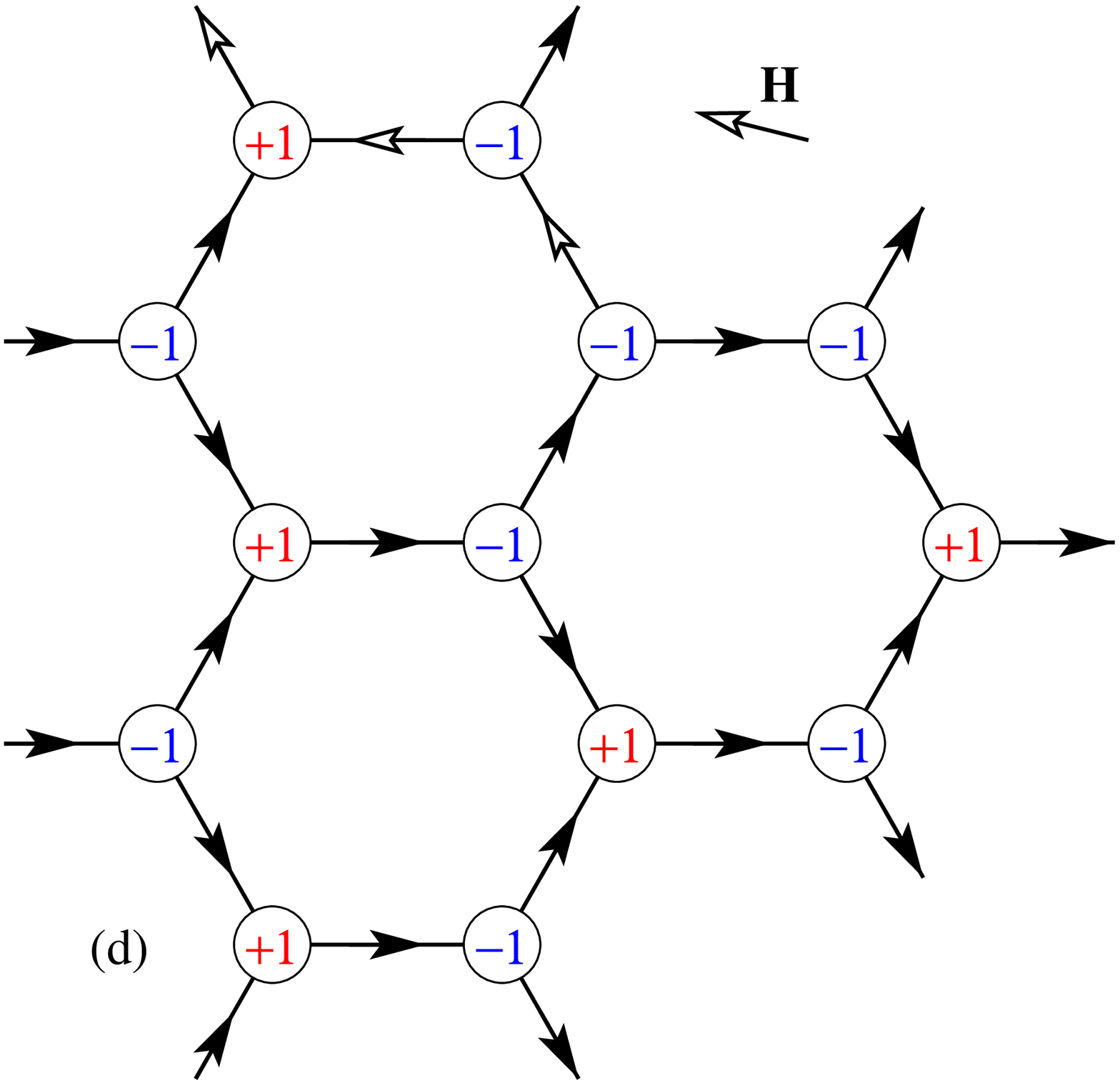}
\end{center}
\caption{Magnetization reversal after the applied field is rotated through $170^\circ$. }
\label{fig:reversal-170}
\end{figure}

When the field is rotated through $\theta = 170^\circ$ relative to the direction of magnetization, the theory described in Section~\ref{sec:numerics-120} no longer applies. Because $H_2$, the reversal field of horizontal links, is lower than $H_1$, these links should reverse first. However, that is impossible because in the initial, fully magnetized state, Figure~\ref{fig:reversal-170}(a), these are minority links whose critical field is roughly $3H_\mathrm{c}$ (Section~\ref{sec:basics-emission}), i.e., much higher than $H_2 = H_\mathrm{c}/\cos{11^\circ} \approx H_\mathrm{c}$.  For this reason, a horizontal link does not reverse until one of its neighbors, a majority link, reverses and in the process alters the charge at one of the horizontal link's ends. This converts the horizontal link into a majority link enabling it to reverse magnetization.  It turns out that this mode of reversal is accompanied by long magnetic avalanches. 

In the simplest scenario, the dynamics begins with the reversal of the weakest link with the critical field near $H_1$, Figure~\ref{fig:reversal-170}(b). The reversal turns the horizontal link next to it into a majority link, which is now ready to reverse since the applied field exceeds its critical field: $H \approx H_1 > H_2$. A $q=-2$ domain wall emitted from its left end travels to the right end where it encounters a site with charge $-1$, Figure~\ref{fig:reversal-170}(b).  As discussed in Section~\ref{sec:basics-absorption}, the arriving domain wall induces the emission of another domain wall into an adjacent link, Figure~\ref{fig:reversal-link}(b). The magnetization of that link gets reversed, bringing us to the state shown in Figure~\ref{fig:reversal-170}(d). The cycle repeats creating an avalanche. In effect, we have a $q=+2$ charge moving along a zigzag path parallel to the applied field and reversing magnetization of the links along the way. The process continues until the moving charge reaches the edge of the system so that an avalanche extends from edge to edge.

\begin{figure}
\includegraphics[width=0.49\columnwidth]{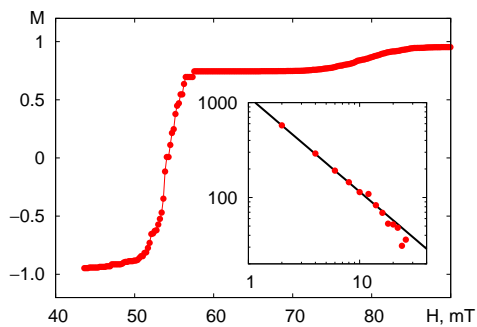}
\caption{Simulated magnetization reversal curve $M(H)$ in an applied field rotated through $170^\circ$. Inset: a log-log plot of the number of avalanches versus their length (red circles) and a fit to the power-law distribution (\ref{eq:1-over-n}) (solid line).}
\label{fig:simulation-170}
\end{figure}

A different scenario may take place if the system has ``weak'' links that trigger the reversal when the applied field is at or below $H_2$. These can be links at the edge of the system or some sort of defects.  Their reversal converts one of the horizontal links (critical field $H_{\mathrm c1}$) to the majority status, as shown in Figure~\ref{fig:reversal-170}(b). When the applied field reaches a value sufficient to induce the reversal of that link, an adjacent link is also reversed as described above, Figure~\ref{fig:reversal-170}(c-d). The next horizontal link down the line (critical field $H_{\mathrm c2}$) will switch immediately if $H_{\mathrm c1} > H_{\mathrm c2}$. The switching will continue until the avalanche comes to a stubborn link whose critical field exceeds $H_{\mathrm c1}$. Its reversal will happen in a higher applied field, possibly triggering another avalanche. 

If the first reversal occurs in a link whose critical field $H_{\mathrm c1}$ is at the lower end of the critical field distribution, the first avalanche will be short because it is unlikely that a large number of subsequent links will have even lower critical fields. As further avalanches get terminated at links with higher critical fields, their lengths will tend to increase. Toward the end of the reversal, avalanches will begin with links whose critical fields are near the higher end of the distribution. These avalanches will be particularly long. The last avalanche in a given string of links will terminate at the edge or will meet an avalanche traveling in the opposite direction. These qualitative considerations anticipate a wide distribution of avalanche lengths. Indeed, we show in \ref{sec:app-avalanches} that the avalanches have a power-law distribution of lengths, 
\begin{equation}
P_n = C/n.
\label{eq:1-over-n}
\end{equation} 
Remarkably, this result applies to any distribution of critical fields, not just a Gaussian one, and numerical simulations confirm this picture. 

As can be seen in Figure~\ref{fig:simulation-170}, magnetization reversal begins in an applied field $H \approx H_2 - \delta H_2 = 47$ mT, where $H_2$ is given by Eq.~(\ref{eq:H1-H2}). At that point, the reversals include single pairs of links from two sublattices. Long avalanches, involving as many as $n = 10$ and more links, are observed by the time the applied field reaches $H \approx H_2 + \delta H_2 = 51$ mT. The length distribution is well fit by the power law (\ref{eq:1-over-n}) as can be seen in the inset of Figure~\ref{fig:simulation-170}. 

The third sublattice reverses in much higher fields, $H \approx H_3 = 77$ mT, where
\begin{equation}
H_3 = H_\mathrm{c}'/\cos{(240^\circ - \theta - \alpha)},
\end{equation}
This stage of the reversal proceeds in the gradual manner described previously.

\section{Discussion}

The dynamics of magnetization in artificial spin ice is a complex problem. In this paper, we have presented a simple model for this system in terms of coarse-grained physical variables (Figure~\ref{fig:variables}), Ising spins $\sigma_{ij}$ living on the links of the spin-ice lattice and magnetic charges $q_i$ residing on its sites. Inspired by our earlier studies of magnetic nanowires \cite{tretiakov:127204, clarke:134412}, where magnetization reversal is mediated by the propagation of domain walls, we have expressed the magnetization dynamics in spin ice in similar terms. Magnetization reversal in individual links of the lattice proceeds through the emission, propagation, and absorption of domain walls with magnetic charge $q_\mathrm{w} = \pm 2$. Coulomb-like interactions between the magnetic charges of the walls and lattice sites play a major role in the dynamics. For example, the magnitude of the critical field, required for the emission of a domain wall, is set by the strength of magnetostatic attraction between a domain wall and the magnetic charge of the lattice site. These heuristic considerations have been confirmed and refined through micromagnetic simulations of a small portion of the spin-ice lattice containing a few links. 

Quenched disorder is another major element affecting the magnetization dynamics. Small imperfections of the artificial lattice are expected to produce a Gaussian distribution of critical fields.  The experimentally measured curve \cite{PhysRevLett.107.167201} is consistent with a Gaussian shape and width $\delta H_\mathrm{c}/\bar H_\mathrm{c} \approx 0.05$. 

The dynamics of magnetization reversal strongly depends on the direction of the external magnetic field. If the field is applied at a small angle relative to the magnetization of a (fully magnetized) sample, $\theta < 131^\circ$ for the parameters we used, the reversal proceeds in a gradual way, with links reversing more or less independently of each other, when the strength of the applied field exceeds the threshold of a given link. For larger angles of rotation, the reversal proceeds in one-dimensional avalanches that can easily span the entire length of the system. The reversal in one link with a critical field $H$ triggers the reversal in several others along the chain. The avalanche stops when it encounters a link whose critical field exceeds $H$. In this regime, avalanche lengths are distributed as a power law, $P_n = C/n$.

It should be pointed out that we model the magnetization dynamics in artificial spin ice as a purely dissipative process, in which the system moves strictly downhill in the energy landscape. Such a picture is very different from an earlier approach extending the notion of an effective temperature to these far-from-equilibrium systems \cite{PhysRevLett.98.217203, PhysRevLett.105.047205}. Whereas energy of a microstate plays a major role in the effective thermal approach, our method puts the focus on energy \textit{gradients}, or forces between magnetic charges. 

This study has a limited scope. We focus on a continuously-connected honeycomb network realized in several experimental studies \cite{Ladak:2010, tanaka:052411, qi:094418} and cover only the basic regimes of its magnetization dynamics, Figure~\ref{fig:regimes}. Interesting phenomena arise at the boundaries between different regimes, particularly when the field is completely reversed, $\theta = 180^\circ$. In this case, avalanches lose their unidirectional character and become random walks. As the magnetization reversal proceeds, avalanches can begin to intersect and block one another. 

Our method can be easily extended to connected networks with other geometries such as square spin ice \cite{nature.439.303}.  Budrikis, Politi and Stamps used a similar heuristic approach to study the dynamics of disconnected magnetic islands \cite{PhysRevLett.105.017201}. 

\section*{Acknowledgments} 

OT and PM thank the Max Planck Institute for the Physics of Complex Systems in Dresden, where part of this work was carried out.  The authors acknowledge support of the Johns Hopkins University under the Provost Undergraduate Research Award (YS) and of the US National Science Foundation under Grants No. DMR-0520491 (OP and PM), DMR-1056974 (SD and JC), and DMR-1104753 (OT).

\appendix

\section{Simulation procedure}
\label{sec:app-procedure}

For a given applied external field, the total magnetic field $\mathbf H$ for each site is computed as a sum of the applied field and the Coulomb fields generated by the charges at the neighboring sites and domain walls (see Section~\ref{sec:basics-disorder}).  For simplicity, we only include the fields from first and second-neighbor sites. Fields of further neighbors decrease rapidly and tend to oscillate in sign. For each link attached to a given site, the program checks whether the net field has a negative projection $H_\mathrm{e} = H \cos{(\theta - \alpha)}$ onto the link's easy axis, Eq.~(\ref{eq:Hc-theta}). If $H_\mathrm{e} < 0$, the program calculates the \emph{weakness} of the site and link, $W = |H_\mathrm{e}| - H_c$. The site and link with the largest $W$ in the sample are considered to be the weakest. As the applied field increases, the largest $W$ becomes positive, triggering the emission of a domain wall from the weakest site into the weakest link. The domain wall propagates to the other end of the link where it is absorbed, either immediately or after the emission of another domain wall as described in Section~\ref{sec:basics}. Once the reversal process that started with the weakest site is complete, the program looks for the next weakest site. The process is repeated until there are no positive $W$ in the system. Spin ice rules are satisfied at each site at all times. No thermal effects are considered.

\section{Statistics of avalanches in the presence of a weak link}
\label{sec:app-avalanches}

Here we derive the statistics of avalanches discussed in Section~\ref{sec:numerics-170}. In this case, the reversal begins in Link 1 (critical field $H$) and spreads to consecutive Links $2, 3, \ldots, n$ (of the same sublattice) as long as their critical fields are lower than $H$. The avalanche stops when it encounters link $n+1$ whose critical field exceeds $H$.  The probability density of the critical-field distribution is $\rho(H)$ and the cumulative probability distribution is 
\begin{equation}
P(H) = \int_{-\infty}^H \rho(H') \mathrm{d}H'.
\end{equation}

Consider an avalanche beginning on link $k$ with a critical field between $H$ and $H + \mathrm{d}H$. The $k-1$ preceding links must have critical fields less than $H$. If the avalanche has length $n$ then links $k+1, k+2, \ldots, k+n-1$ must have critical fields less than $H$, whereas link $k+n$ must have a higher critical field. The probability of such a distribution is 
\begin{equation}
f_n^k(H) \, \mathrm{d}H = [P(H)]^{k-1} \, \rho(H) \, \mathrm{d}H \, [P(H)]^{n-1} \, [1-P(H)].
\end{equation}
However, if the avalanche terminates on link $L$, the last link of the chain, the factor $1-P(H)$ drops out because there is no link $L+1$: 
\begin{equation}
f_n^{L-n+1}(H) \, \mathrm{d}H = [P(H)]^{L-n} \, \rho(H) \, \mathrm{d}H \, [P(H)]^{n-1}.
\end{equation}
The probability to find an avalanche of length $n$ is found by summing this distribution over the initial position of the avalanche $k$ and integrating over the critical field $H$. Performing the sum first, we find
\begin{equation}
f_n = \sum_{k = 1}^{L-n+1} f_n^k = \rho P^{L-1} + \sum_{k=1}^{L-n} \rho (P^{k+n-2} - P^{k+n-1})  
= \rho P^{n-1}.
\end{equation}
The integration of the resulting expression yields the expected number of avalanches of length $n$,
\begin{eqnarray}
F_n &=& \int_{-\infty}^\infty f_n(H) \, \mathrm{d}H
\nonumber\\
	&=& \int_{-\infty}^\infty [P(H)]^{n-1} \rho(H) \, \mathrm{d}H 
	= \int_0^1 P^{n-1} \, \mathrm{d}P = 1/n.
\end{eqnarray}
Note that $F_n$ is an expectation number of avalanches, not a probability distribution normalized to 1. Observing an avalanche of length $n$ does not exclude the possibility of observing an avalanche of a different length $n'$ in the same chain during the same reversal process. The probability that an avalanche will have length $n$ is 
\begin{equation}
P_n = F_n/\sum_{n=1}^L F_n \sim 1/(n \ln{L})
\end{equation}
for large $L$.

\bibliographystyle{iopart-num}
\bibliography{spinice,micromagnetics}

\end{document}